\documentclass[12pt,halfline,a4paper]{pythia}

\begin{document}
	
	\title{Editorial: Understanding Cryptocurrencies*\vspace{0.5cm}}
	
	\author{\name{\large{Wolfgang Karl H\"ardle}}
		\footnotesize{\address{Humboldt-Universit\"at zu Berlin, Germany. \\ Wang Yanan Institute for Studies in Economics, Xiamen University, China.  \\ Sim Kee Boon Institute for Financial Economics, Singapore Management University, Singapore. \\ Faculty of Mathematics and Physics, Charles University, Czech Republic. }
			\email{haerdle[at]hu-berlin.de}}\vspace{-0.5cm}
		\and
		\name\large{{Campbell R. Harvey}}
		\address\footnotesize{{Duke University, Durham, NC, USA.\\ National Bureau of Economic Research, Cambridge MA, USA.}\\
			\email{cam.harvey[at]duke.edu}}\vspace{-0.5cm}
		\and
		\name\large{{Raphael C. G. Reule}}
		\address\footnotesize{{Humboldt-Universit\"at zu Berlin, Germany.}\\
			\email{irtg1792.wiwi[at]hu-berlin.de}}}

	\abstract{Cryptocurrency refers to a type of digital asset that uses distributed ledger, or blockchain, technology to enable a secure transaction. Although the technology is widely misunderstood, many central banks are considering launching their own national cryptocurrency. In contrast to most data in financial economics, detailed data on the history of every transaction in the cryptocurrency complex are freely available. Furthermore, empirically-oriented research is only now beginning, present\-ing an extraordinary research opportunity for academia. We provide some insights into the mechanics of cryptocurrencies, describing summary statistics and focusing on potential future research avenues in financial economics.\\
		
		\footnotesize{
			\vspace{1.5cm}  
			
			$The$ $financial$ $support$ $of$ $Czech$ $Science$ $Foundation$ $under$ $grant$ $no.$ $19-28231X$ $is$ $acknowledged.$ \\

			*Financial support from the Deutsche Forschungsgemeinschaft via CRC 649 ”Economic Risk” and IRTG 1792 ”High Dimensional Non 
			Stationary Time Series” and the European Union’s Horizon 2020
			research and innovation program ''FIN-TECH: A Financial supervision and Technology compliance training programme'' under the grant 
			agreement No 825215 (Topic: ICT-35-2018, Type of action: CSA), Humboldt- Universität zu Berlin, is grate-fully acknowledged.	
}
		
	}\vspace{1cm}

	\date{\today}

	\maketitle

	\newpage


	\normalsize
	
	\section{Introduction}
	\label{sec1}
	
In 2008, the pseudonymous ``Satoshi Nakamoto" posted a white paper describing an implementation of a digital currency called bitcoin that used blockchain technology. More than ten years later, hundreds of cryptocurrencies and innumerable other applications of blockchain technology are readily available.\\
  
 The rise of cryptocurrencies poses an existential threat to many traditional functions in finance. Cryptocurrencies embrace a peer-to-peer mechanism and effectively eliminate the ``middle man", which could be a financial institution. For example, no bank account or credit card is needed to transact in the world of cryptocurrencies. Indeed, a cryptocurrency ``wallet" serves the same function as a bank vault. With a smart phone and the internet, the potential exists for a revolution in financial inclusion --- given that over two billion people are unbanked (GlobalFindex, 2017; World Bank, 2017).\\
  
  The technology, however, goes well beyond providing banking services to the unbanked. It holds the potential for cheap, secure, and near-instant transactions, allowing billions of people to join the world of internet commerce, paying, and being paid, for goods or services, outside of the traditional banking and credit card infrastructure.\\
  
  Cryptocurrencies transactions potentially enable near real-time micropayments. Credit cards are not designed to be used for a one-cent charge to download, for example, a product or service from the internet. Cryptocurrency systems promise to make micropay\-ments seamless and allow businesses to offer real-time pay-per-use consumption of their products, such as video, audio, cell phone service, utilities, and so forth.\\
  
 A cryptocurrency like bitcoin can be thought of as a decentralized autonomous organi\-zation (DAO), an open-source peer-to-peer digital network that enforces the rules it is set up with. In this DAO setting, the money supply is set by an algorithmic rule, and the integrity of the network replaces the need to trust the integrity of human participants. The growth of crypotcurrency technology therefore poses a challenge to traditional monetary authorities and central banks, as Facebook's ``Libra'' coin pre-emission market acceptance suggests (Taskinsoy, 2019). Central banks understand this, and many banks have initiated their own national cryptocurrency initiatives (Bech and Garratt, 2017).\\
  
  As with any new technology, risks are present. In the nascent cryptocurrency market, one concern involves the anonymous nature of transactions in some cryptocurrencies, which could allow nefarious actors to conduct illegal business, or worse, to pose a broader threat to our society and institutions (Foley et al., 2018). The benefits, such as low transaction cost, security and the promise of quick processing, are readily measurable, but quantifying the risks is less straightforward. \\
  
  In our view, any new technology involves risks; if we require no risk, innovation is constrained (Catalini and Gans, 2016). Cryptocurrencies have, in contrast to many markets, a plethora of available and free data, ripe for empirical investigation. We are just now seeing the genesis of academic research focusing on this emerging technology (Harvey, 2014, 2017a, 2017b; Härdle et al., 2018; Kim et al., 2019).\\
  
  We have four goals in this paper. First, we explain the mechanics of cryptocurrencies at a high level. Second, we detail useful data sources for researchers. Third, we provide basic summary statistics given the available data. Finally, we offer a list of possible research applications.

	
	\section{Cryptocurrencies and Blockchains}
	\label{sec2}
	
	The concept of supplementary (Delmolino et al., 2016), alternative (Ametrano, 2016), or digital currencies (Chaum, 1983) is not new, but the concept of an open-source currency without a central point of trust, such as a central distribution agency or state lead control, is new (King and Nadal, 2012). A cryptocurrency is a digital asset designed to work as a medium of exchange using cryptography to secure transactions, to control the creation of additional value units, and to verify the transfer of assets. Many different cryptocurrencies exist, each with their own set of rules, see, for example, \url{coinmarketcap.com} (Iwamura et al., 2014; Abraham et al., 2016; Bartos, 2015; Park et al., 2015). Differences among the cryptocurrencies may involve, for example, the choice of the consensus mechanism, the latency, or the cryptographic hashing algorithms.

	
	\subsection{High-level description of blockchain}
	\label{sec2.1}	

	Abadi and Brunnermeier (2018) describe a blockchain trilemma, i.e. that no ledger can satisfy all ideal qualities of any recordkeeping system --- correctness, decentralization, and  cost efficiency --- simultaneously. Yet, a blockchain is more efficient than a centrally managed traditional ledger (Babich and Hilary, 2018a). A blockchain can be implemented in many ways, but most share several common features. We can think of a blockchain as a very special database. A blockchain’s structure is shared, or $distributed$, rather than centralized, and thus is often referred to as distributed ledger technology (DLT). Figure 1 shows a distributed network. As we discuss later, the distributed network provides some level of security, because it is unlikely an attack can be launched on every copy of the database. Distributed databases are not new, and most distributed databases are not blockchains. The key difference between a regular distributed database and one set on a blockchain is the structure (Babich and Hillary, 2018b).\\
	
	A blockchain is divided into subsheets of data, each one called a $block$. At the end of each block is a $digest$ that summarizes the contents of the block. The digest is repeated as the first line of the next block. If any change is made in the content of a historical block, the digest changes for that block and it will not match the first line of the next block. When the network detects such an inconsistency, it throws out the corrupted block and replaces the block with the original. In this sense, the database is immutable. Given this structure (i.e., data organized in blocks with updates to the blockchain being append-only, based on the respective consensus mechanism), it is extremely unlikely that history can be rewritten. The digest at the end of a block and at the beginning of the next is generated by a cryptographic hashing function.
	
	\vspace{-0.5cm}
	
	\begin{figure}[H]
		\hfill
		\subfigure[Centralized]{\includegraphics[width=4cm]{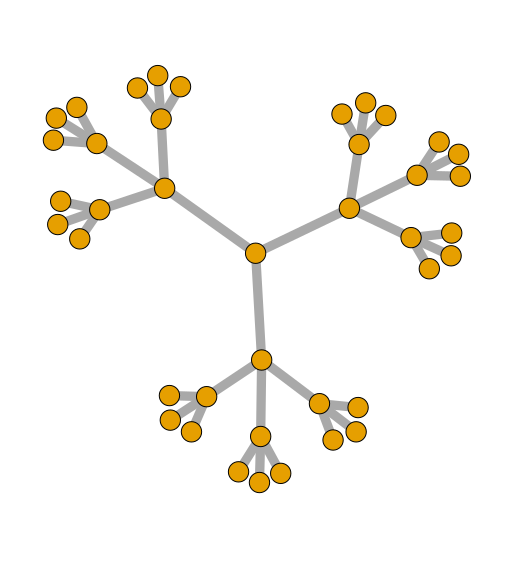}}
		\hfill
		\subfigure[Decentralized]{\includegraphics[width=4cm]{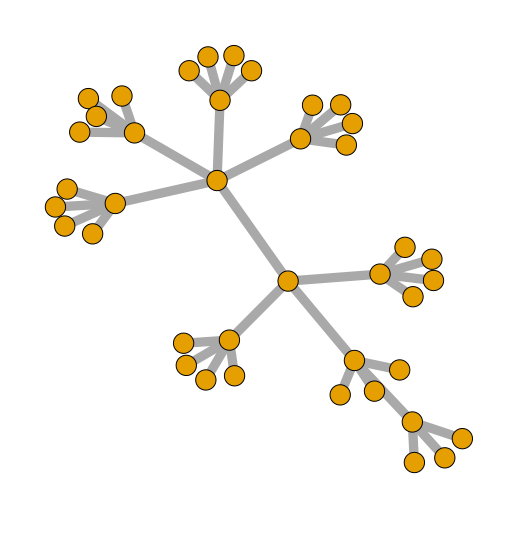}}
		\hfill
		\subfigure[Distributed]{\includegraphics[width=4cm]{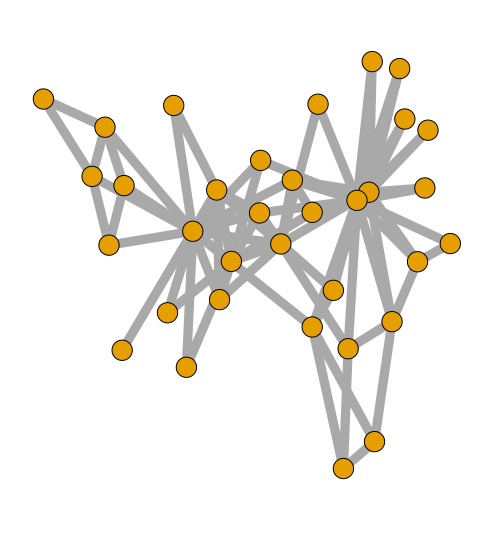}}
		\hfill
		\href{http://github.com/QuantLet/CrixToDate}{\includegraphics[keepaspectratio,width=0.4cm]{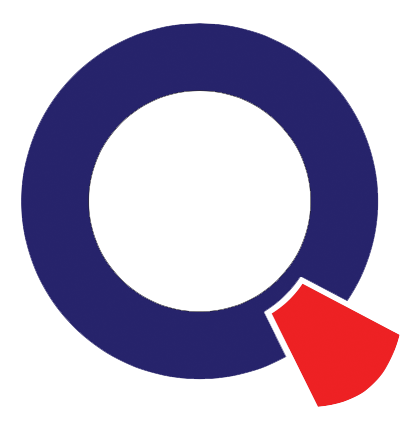}}
		\caption{Types of networks. (github.com/QuantLet/CrixToDate)}
	\end{figure}

		\vspace{-0.1cm}

	All presented numerical and pictoral examples shown are reproducible and can be found on  \includegraphics[keepaspectratio,width=0.4cm]{media/qletlogo_tr.png} \url{www.quantlet.de} (Borke and H\"ardle, 2018).

	
	\subsection{Hashing}
	\label{sec2.2}	
	
	A hash function is a one-way mathematical algorithm that takes an input and transforms it into an output, known as the $hash$ or $digest$. Hashing functions have a long history in computer science and are integral to the blockchain technology. Hashing should not be confused with encryption. With encryption, a file is encrypted with a key and decrypted with a key. Hashing has no decryption step. Additionally, a good hashing algorithm makes it computationally infeasible to find two input values that produce the same hash value (output); this is known as $collision$ $resistance$ (Paar and Pelzl, 2010; Derose, 2015; Harvey, 2016). \\
	
	One common cryptographic hashing algorithm, the Secure Hash Algorithm (SHA-256), has a maximum input size of 2\textsuperscript{64}‐1 bits (more than 2 million terabytes) and an output of 256 bits. We usually represent the SHA-256 output in hexadecimal form, also called base 16 (the characters 0-9 and a-f). In order to make the theoretical maximal input size more visual, we assume that 1 bit equals to 1 mm\textsuperscript{2}. A soccer field has the dimensions of 7,140 m\textsuperscript{2}, therefore 2\textsuperscript{64}‐1 bits could theoretically fill 2,583,577,601 soccer fields. As the whole surface of the earth equals to 510,000,000,000,000 m\textsuperscript{2}, we could also cover around 36170 times the earth with the theoretical input size. This input information will be stored in a very short output, the hash. If just one piece of the input, like for example a blank space or a comma, is changed, then the hash output be completely different.\\
	
	Importantly, the digest does not reveal the original information. For example, suppose we want to send an electronic document via email, but are worried that the document could be corrupted and the content altered. One way to verify the integrity of the email is to use a hashing function, such as the SHA-256. Before sending the email, we obtain a SHA-256 of the document and post the SHA-256 on our website. We then send the document. The recipient also hashes the document to verify the hash is the same as the hash on our website. If they are identical, we have securely sent the document. Posting the hash on our website does not reveal the content of the email.\\

		\begin{raggedright}
	Here are some examples, which you can try by using the R package \href{https://cran.r-project.org/web/packages/digest/digest.pdf}{``digest"} (cran.r-pro ject.org/web/packages/digest), the Python libary \href{https://docs.python.org/2/library/hashlib.html}{``hashlib"} (docs.python.org/2/library/ hashlib.html), or many online programs such as \href{http://emn178.github.io/online-tools/sha256.html}{interactive Github-based repositories} (emn178.github.io/online-tools/sha256.htm).\vspace{0.4cm}

	\textbf{Input:} H\textbf{e}llo CRIX\\
	\textbf{Output:} f9a2b57d86cc4ba463a3bedbbe0c7e850da5b34c6bcc1a92b794308ceaf93761
	\vspace{0.2cm}
	
	\textbf{Input:} H\textbf{a}llo CRIX\\
	\textbf{Output:} 0198c2ea3632efd2758cd40a5609037fe4aa1590850339ad6d6a7fd3e518ec65\\

\vspace{0.5cm}
Note that changing a single letter from ``e" to ``a" completely changes the hash.
	
	\end{raggedright}
	
	
	\subsection{Blockchains}
	\label{sec2.3}
	
	Haber and Stornetta (1991) were the first to propose a linear hash chain or blockchain. They solved the problem of how to certify when a digital document was created or last changed by timestamping a cryptographic hash of the document. By not timestamping the data itself, the privacy of the content was preserved. Haber and Stornetta’s time-stamping proposal also solved the potential problems of collusion and lack of trust by linking hash values together and using digital signatures, which uniquely identify the signer.\\
	
A year later, Dwork and Naor (1992) proposed a proof-of-work system to combat junk email. Their idea was to provide each email with a header containing virtual postage in the form of a single calculation, which the receiver could verify with very little effort. This postage stamp was to be proof that a modest amount of CPU time was expended for calculating the stamp prior to sending the email. Whereas an individual email could be sent at a very low cost, the intent was to defeat spammers, who send millions of emails. Spamming would come at a high price. Back (2002) coined the term $hashcash$ to describe this $proof$ $of$ $work$, the computational cost of producing each hash, a term first used by Jakobsson and Juels (1999).\\
	
Many applications of blockchain technology exist, but we focus our attention on cryptocurrencies. \href{https://bitcoin.org/en/}{Bitcoin} (cryptocurrency known as BTC; bitcoin.org) was the first example of a digital asset, which has no backing or intrinsic value, based on blockchain technology (Nakamoto, 2008; Böhme et al., 2015, for a review).\\
	
	A common characteristic of cryptocurrencies is a network of peers with equal standing. Each participant has a copy of the ledger and offers an algorithmic consent on the correct ledger (i.e., which new block is accepted and which block is rejected to form a new part of the blockchain). It is unneccesary to know your peers in a blockchain or to trust them. It is also possible to design a blockchain so that only specific trusted parties have the ability to add to the ledger. Private, permissioned blockchains are a source of considerable interest for many central banks (MAS, 2017; Bundesbank, 2017; SARB, 2018). In contrast to cryptocurrencies such as bitcoin, trust is necessary in the permissioned blockchain, because the central banks actually ``own" the coins, i.e. as a governing layer they have the right to change the supply of coins (Bordo and Levin, 2017). \\
	
	 Any type of transaction, for example, a financial contract for any type of property transfer, can be put into a blockchain. Given its immutability, a blockchain provides an official record of the contract and a single agreed-upon version of the contract, which is unlikely to be disputed.\\
	
	To summarize, a blockchain is distinguished from an ordinary distributed database by its unique structure, which linearly connects smaller pieces of the database, or the blocks. The chaining comes in the form of a cryptographic hashing function. Any change to history will break the chain on a particular copy of the database. When a chain is broken, the network fixes it by replacing any corrupted block with a valid block.

	
	\subsection{Cryptocurrencies}
	\label{sec2.4}
	
A currency without an intrinsic value, such as a cryptocurrency like bitcoin, can only function if sufficient market acceptance is present and if the belief exists that the currency has the value attributed to it. With a conventional fiat system, money has value because people trust the central bank. For a cryptocurrency, additions to the public ledger are confirmed by a crowd of participants. There is no central bank and participants do not need to trust each other --- trust only applies to the algorithm and the network that defines the particular blockchain. A transaction is only valid if the output is equal to the input, that is, the transactor actually has the funds she or he wants to transfer. The only exceptions are new issues of the cryptocurrency, which are algorithmicly predetermined.\\

We have demonstrated the simplicity of creating a SHA-256 hash to link one block to the next. Why is it then that massive computing power is needed to maintain the bitcoin network? The power required has to do with the proof-of-work consensus concept. The danger of using a simple SHA-256 is that a nefarious actor could change a historical block and all subsequent blocks, essentially rewriting history, by ensuring all hashes match. To make this unlikely, Nakamoto (2008) proposed the idea of requiring ``work". Thus, instead of simply providing any SHA-256 output, a special SHA-256 output, which has many leading zeros, is required. In other words, the proposed SHA-256 hash needs to be lower than or equal to the current $target$ in order for the block to be accepted by the network as the next block to be added to the blockchain. This ``difficulty" ensures that a new block is added on average every \href{https://bitinfocharts.com/comparison/bitcoin-confirmationtime.html}{10 minutes} (bitinfocharts.com/comparison/bitcoin-confirmationtime.html)to the bitcoin blockchain (so-called block time). To find this special hash, certain nodes, called $miners$, will take a candidate group of verified transactions and cycle through numbers, say, 1, 2, 3, … [very large number], until the output of the SHA-256 has some leading zeros. This number, which is added to a digest of the transactions, is called a $nonce$.\\

The computing power requirement arises because the leading zeros are determined via a brute-force search. The probability of one leading zero is 1/16, but the probability of, for example, 18 leading zeros is a very small number, $(1/16)^{18}$. The search is why the vast computing power is needed, see subsection 2.2.\\
 
The first miner that finds the (currently) 18 leading zeros, as in our example, presents its group of transactions and the nonce to the network. Verifying that the transactions plus nonce delivers the leading zeros is easy. Once each node verifies the candidate block, the new block is added to the bitcoin blockchain. This process is the bitcoin consensus mechanism. The miner that found the winning block is rewarded with freshly ``minted" bitcoin. If technology advances or additional computing joins the mining efforts in the network so that blocks are being solved in less than 10 mintues, the algorithm adjusts the difficulty to, perhaps, 19 leading zeros. If computing power leaves the network, the difficulty can be reduced.\\

Cryptocurrency mining is therefore analogous to gold mining. Gold mining is expensive. Cryptocurrency miners spend computing power to find the hash as described above. A gold miner only gets rewarded if gold is found. Cryptocurrency miners only get rewarded if they are the first to find the winning hash. Like mining for gold, mining for cryptocurrency is risky. The continuous expenditure of resources such as for hardware and energy (see also subsection 4.9) for a prolonged period without being rewarded is an inherent risk.\\

Proof of work makes it unlikely that a historical block and all subsequent blocks can be altered, but securing the highly specialized computing power needed to rewrite history is not currently likely. Nakamoto (2008) states that if a single entity gains 51\% of the computing power, it is possible. \\
 
Proof of work is only one approach to consensus, many alternative mechanisms exist and they may not entail the high equipment and energy costs that bitcoin miners face. The second leading cryptocurrency, \href{https://www.ethereum.org/}{Ethereum} (ethereum.org), uses a similar proof-of-work mechanism. Ethereum, however, has committed to change to a proof-of-stake mechanism (Franco, 2015; ETH, 2018). Instead of allocating block mining proportionally to the relative hashing power, the proof-of-stake protocol allocates blocks proportionally to the current holdings (Buterin, 2014; Cotillard, 2015). As a result, the participants with the most cryptocurrency are particularly incented to do the right thing to keep the system running and healthy. Such a method holds the promise of much-improved latency and substantially less energy consumption. A participant who possesses 1\% of the cryptocurrency could mine 1\%, on average, of the proof-of-stake blocks. Ethereum has a number of other differences from bitcoin. Ethereum blocks are added approximately every  \href{https://etherscan.io/chart/blocktime}{14 seconds} (etherscan.io/chart/blocktime) rather than every 10 minutes, and importantly, ethereum allows for $smart$ $contracts$, or small computer programs, to be deployed in its blockchain. These smart contracts are run redundantly on each node.\\ 
    
     Many other consensus mechanisms are currently available: STEEM's $proof$ $of$ $brain$ rewards participants for creating and curating content in their social network (\href{https://steem.io/steem-bluepaper.pdf}{STEEM.io Bluepaper}; steem.io/steem-bluepaper.pdf) and Slimcoin's $proof$ $of$ $burn$ bootstraps one cryptocurrency off another by demonstrating proof of having ``burnt" some units of value by sending a specific amount to a verifiable unspendable address (\href{https://github.com/slimcoin-project/slimcoin-project.github.io/raw/master/whitepaperSLM.pdf}{Slimcoin Whitepaper}; github.com/slimcoin-project/slimcoin-project.github.io/raw/master/whitepaperSLM.pdf), or different implementations of the Byzantine fault tolerance, which was first described as the Byzantine Generals' Problem by Lamport et al. (1982), are used by systems such as \href{https://neo.org}{NEO} (neo.org), \href{https://stellar.org}{Stellar} (stellar.org) and  \href{https://hyperledger-fabric.readthedocs.io/}{Hyperledger Fabric} (hyperledger-fabric.readthedo cs.io).

	
	\subsection{Not all cryptocurrencies are the same}
	\label{sec2.5}

	We can group cryptocurrencies into seven broad classes. Bitcoin falls into the first category; it was originally designed as a $transaction$ $mechanism$. Think of it as Gold 2.0.   \href{https://litecoin.org/}{Litecoin} (litecoin.org) is very similar to bitcoin and was one of the first alternatives to bitcoin. Litecoin’s blocks are added every \href{https://bitinfocharts.com/comparison/litecoin-confirmationtime.html}{2.5 minutes} (bitinfocharts.com/comparison/lite coin-confirmationtime.html), on average, compared to every \href{https://bitinfocharts.com/comparison/bitcoin-confirmationtime.html}{10 minutes} for bitcoin.\\
	
	Ethereum falls into the second class: a $distributed$ $computation$ token. As mentioned earlier, it is possible to run a computer program on the ethereum network. Think of it as an Internet computer where small programs, smart contracts, are executed when called upon, on every node. Other examples in this class include \href{https://tezos.com/}{Tezos} (tezos.com), \href{https://eos.io/}{EOS} (eos.io) and \href{https://dfinity.org/}{DFinity} (dfinity.org).\\ 

The third class of cryptocurrency is called a $utility$ token. A utility token is a programmable blockchain asset. One example is \href{https://golem.network/}{Golem} (golem.network), a currency that allows the user to buy computing power from a network of users or to sell excess capacity to others. \href{https://storj.io/}{Storj} (storj.io)is similar and allows the user to rent out unused disk storage. Other examples in this class are \href{https://sia.tech/}{Sia} (sia.tech) and \href{https://filecoin.io/}{FileCoin} (filecoin.io). \\

The fourth class of cryptocurrency is a $security$ token, a token that represents stocks, bonds, derivatives, or other financial	assets. New security token offerings are called STOs. This type of token could lead to substantial efficiency gains in both clearing and settlement.\\

The fifth class is called $fungible$ tokens. The most popular is called ERC-20 which is issued on the ethereum blockchain. Here a small amount of ETH represents something different – and more valuable.\\

A $non$-$fungible$ token is the sixth classification. In this case, each token is unique and not interchangeable with another. One popular protocol is ethereum’s ERC-721. Dhrama debt agreements fall into this classification. Two other  eamples of non-fungible tokens are  \href{https://www.cryptokitties.co/}{Cryptokitties} (cryptokitties.co) and  \href{https://decentraland.org/}{Decentraland (LAND; decentraland.org)}.\\

The final class of cryptocurrencies are called $stablecoins$. There are four categories. The first category is collateralized with fiat currency. This includes stablecoins such as \href{https://tether.to/}{tether (USDT)}(tether.to)  and \href{https://www.circle.com/en/usdc}{Circle’s USDC} (circle.com). These cryptocurrencies are designed to be fully collaterized by US dollar deposits. \href{https://lbx.com/blog/lbx-peg/}{LBXPeg} (lbx.com/blog/lbx-peg) is tied to pound sterling. An emerging market, Mongolia has a cryptocurrency called \href{https://www.candy.mn/#/}{Candy} (candy.mn) tied to their currency. This class also includes national cryptofiats. As mentioned earlier, many central bank are investigating the potential  \href{https://law.yale.edu/system/files/area/center/global/document/411_final_paper_-_fedcoin.pdf}{Fedcoin} (US Federal Reserve),  \href{https://coinmarketcap.com/currencies/eurocoin/}{Eurocoin (European Central Bank)},  \href{https://www.r3.com/wp-content/uploads/2017/06/cadcoin-versus-fedcoin_R3.pdf}{CADCoin} (Bank of Canada), for example. Venezuela already issued a national crypto called \href{https://www.petro.gob.ve/}{Petro} (petro.gob.ve).\\

The second category of stablecoins are collateralized with real assets.  Examples include currencies that are collateralized by gold (\href{https://digix.global/}{Digix Gold, DGX}; digix.global), a basket of seven precious metals used in technology (\href{https://www.tiberiuscoin.com/}{Tiberius coin, TCX}; tiberiuscoin.com) or even Swiss real estate (\href{https://www.swissrealcoin.io/}{Swiss Real Coin, SRC}; swissrealcoin.io).\\

The third category of stablecoins are cryptocurrency collateralized. The leading example is the collateralized debt positions that \href{https://makerdao.com/en/}{MakerDAO} offers that enable their \href{https://makerdao.com/en/dai}{DAI coin} (makerdao.com/en/dai) to be pegged to the US dollar.\\

The final category of stablecoins are uncollateralized. An example of this type of inititive is the \href{https://www.basis.io/}{Basis project} (basis.io) and their basecoin which has been put on hold given regulatory concerns.\\

This list of classifications is not exhaustive because many cryptocurrency concepts, such as \href{https://overlay.market}{Overlay} (overlay.market) or \href{https://developers.libra.org/docs/assets/papers/libra-move-a-language-with-programmable-resources.pdf}{Facebook's Libra} (libra.org), do not  easily fit within our seven-category taxonomy. Our point is simple: cryptocurrencies have many uses and characteristics that extend beyond the traditional cryptocurrencies of bitcoin and ethereum.

	
	\section{Summary Analysis of Cryptocurrencies}
	\label{sec3}
	
	We will now focus on an econometric analysis of the currently most liquid cryptocurrencies. Valuation of currencies that are not collateralized or linked to real assets is a challenge. These currencies are highly volatile and subject to bubble-like behavior. These currencies, however, provide an ideal testing ground for economic theory. In the fall of 2017, bitcoin rose to over \$19,000. The bubble burst in 2018. Because every bitcoin transaction is freely available, we are provided with an extraordinary research opportunity. We begin with a simple benchmarking analysis using the S\&P 500 Index (S\&P 500), SPDR Gold Shares (GOLD), and CBOE Volatility Index (\href{http://www.cboe.com/vix}{VIX}; cboe.com/vix), which measures the implied volatility of the S\&P 500 index.

	
	\subsection{Cryptocurrency data sources}
	\label{sec3.1}

Many, sometimes very generic, data sources are available for cryptocurrencies, which unlike traditional assets trade 24/7, creating a vast amount of data to capture. Blockchain-based systems --- most of which are open to the public for participation --- have data that are readily available using basic API's (application programming interface).\footnote{For this brief analysis, we are using cryptocurrency data provided by the \href{http://thecrix.de/}{CRIX database} \href{http://thecrix.de/}{(thecrix.de)} and the \href{https://www.cryptocompare.com/api/}{Cryptocompare API} (cryptocompare.com/api), as well as data for the traditional assets provided through the Bloomberg Terminal. Further information can also be found at \href{https://lopp.net/bitcoin.html}{Jameson Lopp's Bitcoin Resources} (lopp.net/bitcoin.html).} In subsection 4.3, we discuss exchange APIs, which provide the data for actual cryptocurrency market transactions (Guo and Li, 2017). Several of the more important data sources include \href{https://www.coingecko.com/en}{CoinGecko} (coingecko.com), a cryptocurrency ranking and evaluation site that breaks down quanti\-tative and qualitative data for a number of different metrics, as well as \href{https://coinmarketcap.com/}{Coinmarketcap} (coinmarketcap.com),  \href{https://onchainfx.com/}{Onchainfx} (onchainfx.com),  \href{https://www.cryptocompare.com/}{Cryptocompare} (crypt ocompare.com),  \href{http://www.bitinfocharts.com}{BitInfoCharts} (bitinfocharts.com),  \href{https://coincheckup.com/}{CoinCheckup} (coincheckup.com), and \href{https://coincodex.com/}{Coincodex} (coincodex.com). Each has unique attributes.

	
	\subsection{Statistical  overview of cryptocurrencies}
	\label{sec3.3}
	
	While there are thousands of cryptocurrencies, we focus our analysis on three: \href{https://bitcoin.org/en/}{bitcoin} (BTC), \href{https://ethereum.org/}{ethereum} (ETH), and \href{http://www.ripple.com}{Ripple} (XRP; ripple.com). To represent traditional assets, we have chosen S\&P 500, GOLD, and VIX. Each cryptocurrency has a different implementation. Some, like Litecoin, are very similar to BTC. As previously mentioned, ETH allows for distributed computation. In contrast to BTC, ETH may be easier to value because it has a tangible component (i.e., running a computer program on a network).\\
	
	XRP focuses on the banking sector with the promise of fast and secure transfers of tokens, whether in fiat, cryptocurrency, commodity, or other unit of value, across different networks, geographic borders, and currencies (Aranda and Zagone, 2015). The Ripple system’s efficiency and security challenges the traditional SWIFT system for transfers, which is now also interested in blockchain-based technologies (Arnold, 2018).\\

	In Figure 2, we show the cumulative return over time for \textbf{BTC}, \textcolor{JungleGreen}{XRP}, \textcolor{Dandelion}{ETH}, \textcolor{red}{SPDR GOLD Shares} and \textcolor{blue}{S\&P 500} from May 1, 2017 to Jun. 30, 2019. We chose this short time period because, prior to this, the cryptocurrency market was substantially illiquid; it was not until 2016 that the initial influx of exchanges and users entered the market. By May 2017, all three cryptocurrencies were active and had achieved  sufficiently high market capitalizations. Although BTC has the highest market capitalization and had received intensive media exposure prior to our sample period, this time period allows us to capture both the liquid trading period and the full sample of ETH and XRP in addition to BTC.\\

	\begin{figure}[H]
		\centering
		\includegraphics[keepaspectratio,width=10cm]{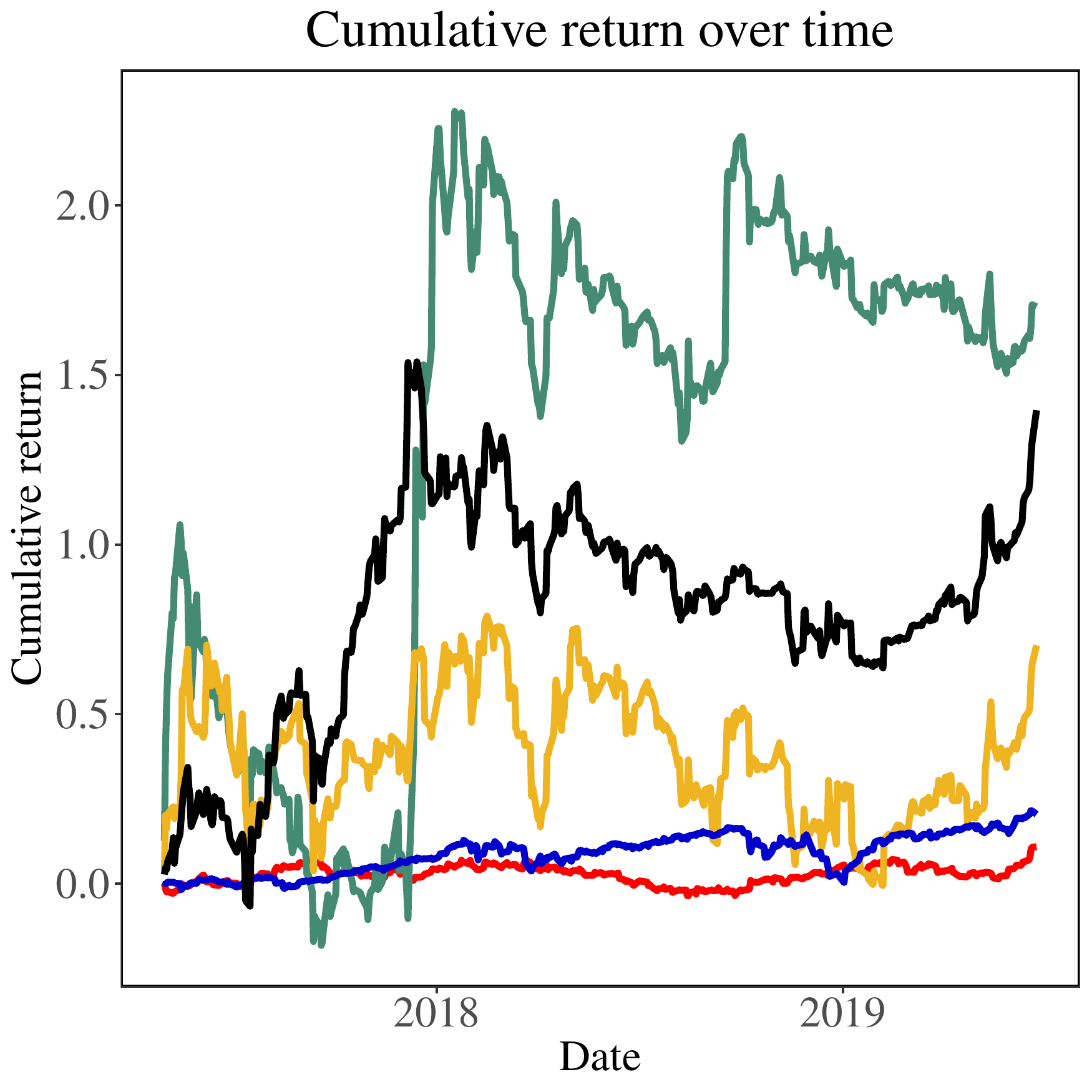}
		
		\vspace{-0.3cm}
		\caption{Cumulative return over time between May 1, 2017 and Jun. 30, 2019 of \textbf{BTC}, \textbf{\textcolor{JungleGreen}{XRP}}, \textbf{\textcolor{Dandelion}{ETH}}, \textbf{\textcolor{red}{GOLD}} and \textbf{\textcolor{blue}{S\&P 500}}. (github.com/QuantLet/UCC)}
	\end{figure}

\vspace{-0.5cm}

\textcolor{white}{XXXXXXXXXXXXXXXXXXXXXXXXXXXXXXXXXXXXXXXXXXXXXLXX}\href{https://github.com/QuantLet/UCC}{\includegraphics[keepaspectratio,width=0.4cm]{media/qletlogo_tr.png}}

Tables 1 and 2, as well as Figure 3, provide the correlations of the daily and monthly returns from May 1, 2017 to Jun. 30, 2019 for the five assets and VIX. Red indicates a positive correlation and blue indicates a negative correlation, with significant correlations being marked in a darker color. The correlations are likely time varying. Figure 3 shows the correlation time series of rolling windows of one trading year (250 days) relative to BTC. Both XRP and ETH are positively correlated with BTC. No evidence is shown of a significant correlation with S\&P 500, GOLD or the VIX.\\


\begin{singlespace}
	
\begin{table}[H]
	\centering
	\caption{Daily Correlation, May 1, 2017 to Jun. 30, 2019.}
\begin{tabular}{rrrrrrrr}
		\hline
		\hline
\cellcolor[HTML]{FFFFFF}Daily & BTC                           & ETH                           & XRP                           & GLD                           & SP500                         & VIX                           \\ \hline
BTC                           &                               & \cellcolor[HTML]{FD6864}0.42  & \cellcolor[HTML]{FFCCC9}0.21  & \cellcolor[HTML]{FFCCC9}0.04  & \cellcolor[HTML]{FFCCC9}0.04  & \cellcolor[HTML]{ECF4FF}-0.06 \\
ETH                           & \cellcolor[HTML]{FD6864}0.42  &                               & \cellcolor[HTML]{FFCCC9}0.20  & \cellcolor[HTML]{FFCCC9}0.06  & \cellcolor[HTML]{FFCCC9}0.01  & \cellcolor[HTML]{ECF4FF}-0.01 \\
XRP                           & \cellcolor[HTML]{FFCCC9}0.21  & \cellcolor[HTML]{FFCCC9}0.20  &                               & \cellcolor[HTML]{FFCCC9}0.04  & \cellcolor[HTML]{ECF4FF}-0.01 & \cellcolor[HTML]{ECF4FF}-0.02 \\
GLD                           & \cellcolor[HTML]{FFCCC9}0.04  & \cellcolor[HTML]{FFCCC9}0.06  & \cellcolor[HTML]{FFCCC9}0.04  &                               & \cellcolor[HTML]{ECF4FF}-0.15 & \cellcolor[HTML]{FFCCC9}0.13  \\
SP500                         & \cellcolor[HTML]{FFCCC9}0.04  & \cellcolor[HTML]{FFCCC9}0.01  & \cellcolor[HTML]{ECF4FF}-0.01 & \cellcolor[HTML]{ECF4FF}-0.15 &                               & \cellcolor[HTML]{71A6F2}-0.80 \\
VIX                           & \cellcolor[HTML]{ECF4FF}-0.06 & \cellcolor[HTML]{ECF4FF}-0.01 & \cellcolor[HTML]{ECF4FF}-0.02 & \cellcolor[HTML]{FFCCC9}0.13  & \cellcolor[HTML]{71A6F2}-0.80 &                               
\\ \hline
\hline
&&&&&&
\href{https://github.com/QuantLet/UCC}{\includegraphics[keepaspectratio,width=0.4cm]{media/qletlogo_tr.png}}
	\end{tabular}
\end{table}

\end{singlespace}

\vspace{-0.5cm}

\begin{singlespace}

\begin{table}[H]
	\centering
		\caption{Monthly Correlation, May 1, 2017 to Jun. 30 2019.}
\begin{tabular}{rrrrrrrrr}
		\hline
\hline
      & BTC                           & ETH                           & XRP                           & GLD                           & SP500                         & VIX                           \\ \hline
BTC   &                               & \cellcolor[HTML]{FD6864}0.48  & \cellcolor[HTML]{FD6864}0.45  & \cellcolor[HTML]{FFCCC9}0.08  & \cellcolor[HTML]{FFCCC9}0.13  & \cellcolor[HTML]{ECF4FF}-0.08 \\
ETH   & \cellcolor[HTML]{FD6864}0.48  &                               & \cellcolor[HTML]{FD6864}0.58  & \cellcolor[HTML]{FFCCC9}0.26  & \cellcolor[HTML]{FFCCC9}0.12  & \cellcolor[HTML]{ECF4FF}-0.19 \\
XRP   & \cellcolor[HTML]{FD6864}0.45  & \cellcolor[HTML]{FD6864}0.58  &                               & \cellcolor[HTML]{FFCCC9}0.15  & \cellcolor[HTML]{ECF4FF}-0.08 & \cellcolor[HTML]{FFCCC9}0.02  \\
GLD   & \cellcolor[HTML]{FFCCC9}0.08  & \cellcolor[HTML]{FFCCC9}0.26  & \cellcolor[HTML]{FFCCC9}0.15  &                               & \cellcolor[HTML]{ECF4FF}-0.10 & \cellcolor[HTML]{FFCCC9}0.17  \\
SP500 & \cellcolor[HTML]{FFCCC9}0.13  & \cellcolor[HTML]{FFCCC9}0.12  & \cellcolor[HTML]{ECF4FF}-0.08 & \cellcolor[HTML]{ECF4FF}-0.10 &                               & \cellcolor[HTML]{71A6F2}-0.75 \\
VIX   & \cellcolor[HTML]{ECF4FF}-0.08 & \cellcolor[HTML]{ECF4FF}-0.19 & \cellcolor[HTML]{FFCCC9}0.02  & \cellcolor[HTML]{FFCCC9}0.17  & \cellcolor[HTML]{71A6F2}-0.75 &                         
                          \\ \hline
\hline
&&&&&&
\href{https://github.com/QuantLet/UCC}{\includegraphics[keepaspectratio,width=0.4cm]{media/qletlogo_tr.png}}
	\end{tabular}
\end{table}

\end{singlespace}

\vspace{-0.5cm}

(github.com/QuantLet/UCC)\\

We note, first, the cryptocurrencies are positively correlated, which is especially evident in an analysis of the monthly data. Second, the correlations of the cryptocurrencies with both S\&P 500 and GOLD are relatively low over the limited sample. We also include the correlation with VIX, which largely hovers around zero.\\

Figure 4 plots the 100-day rolling window standard deviations for each asset and VIX. Most cryptocurrencies are an extremely risky store of value given their volatility, which is evident from the volatility of the cryptocurrencies being much higher than those of GOLD and S\&P 500.

\begin{figure}[H]
		\centering
		\includegraphics[keepaspectratio,width=10.5cm]{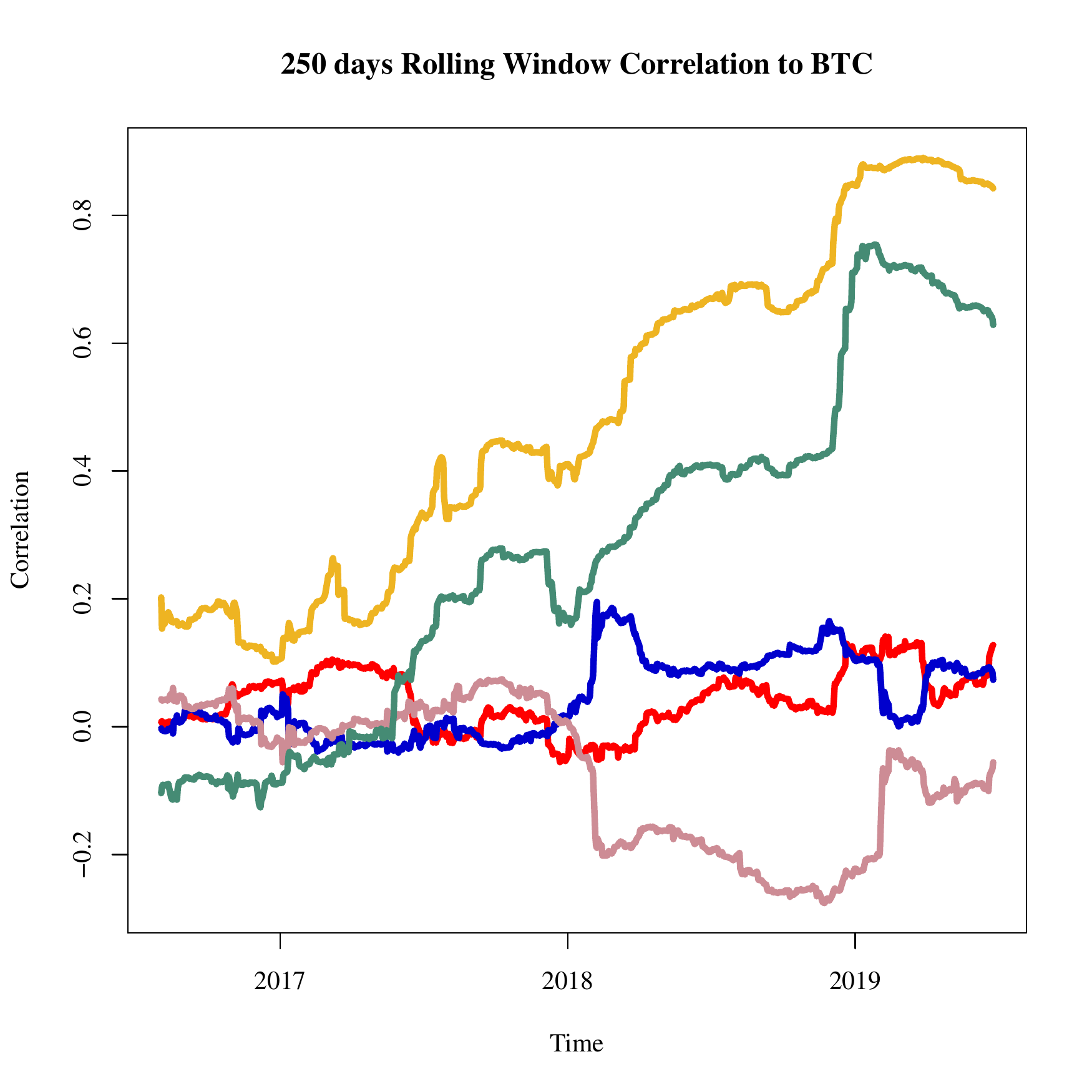}
		
		\vspace{-0.3cm}
		\caption{250 days Rolling Windows Correlations of  \textbf{\textcolor{JungleGreen}{XRP}}, \textbf{\textcolor{Dandelion}{ETH}}, \textbf{\textcolor{red}{GOLD}}, \textbf{\textcolor{blue}{S\&P 500}} and \textbf{\textcolor{Tan}{VIX}} to \textbf{BTC}; daily data, May 1, 2017 to Jun. 30, 2019. (github.com/QuantLet/UCC)}
\end{figure}

\vspace{-0.5cm}

\textcolor{white}{XXXXXXXXXXXXXXXXXXXXXXXXXXXXXXXXXXXXXXXXXXXXXLXX}\href{https://github.com/QuantLet/UCC}{\includegraphics[keepaspectratio,width=0.4cm]{media/qletlogo_tr.png}}

\vspace{-0.7cm}

\begin{figure}[H]
	\centering
	\includegraphics[keepaspectratio,width=10.5cm]{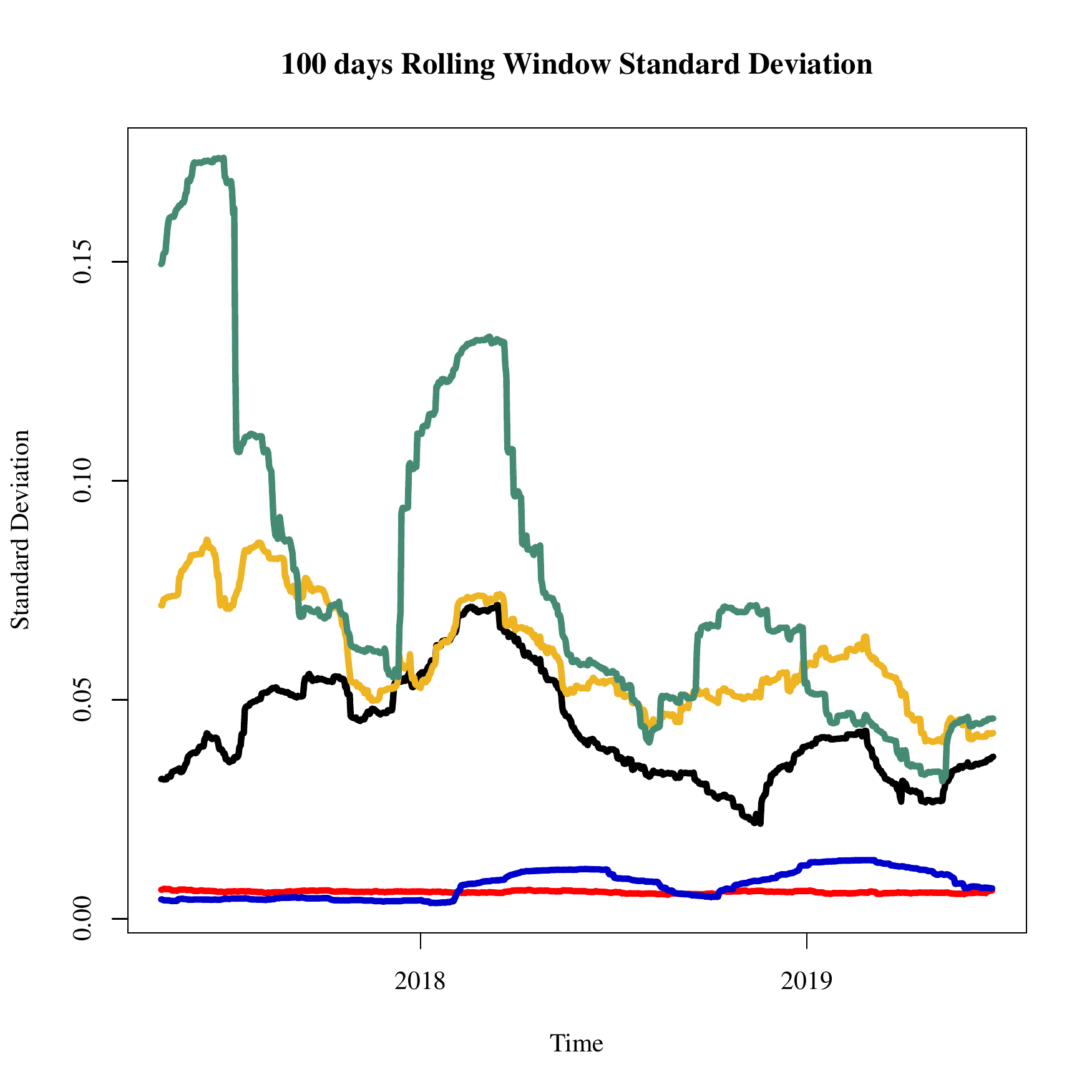}
	\caption{100 days Rolling Window Standard Deviation of \textbf{BTC},  \textbf{\textcolor{JungleGreen}{XRP}}, \textbf{\textcolor{Dandelion}{ETH}}, \textbf{\textcolor{red}{GOLD}} and \textbf{\textcolor{blue}{S\&P 500}}; daily data, May 1, 2017 to Jun. 30, 2019. (github.com/QuantLet/UCC)}
\end{figure}

\vspace{-0.5cm}

\textcolor{white}{XXXXXXXXXXXXXXXXXXXXXXXXXXXXXXXXXXXXXXXXXXXXXLXX}\href{https://github.com/QuantLet/UCC}{\includegraphics[keepaspectratio,width=0.4cm]{media/qletlogo_tr.png}}

	Further insights into the distributional properties of cryptocurrencies can be gained by studying the higher moments of returns, for example, excess kurtosis and skewness, as shown in Table 3. Not surprisingly, the higher moments of the cryptocurrencies are far from what we would expect for a normal distribution. This observation is also evident in the QQ plots for BTC and GOLD shown in Figure 5.\\
		
		
	\begin{table}[H]
		\centering
		\caption{Log Daily Returns Statistics, May 1, 2017 to Jun. 30, 2019.}
		\begin{tabular}{lrrrrrrr}
			\hline
			\hline
			                   & Mean                       & Std. Dev.                  & Skewness                   & e. Kurtosis                    & Min.                       & Max.                       \\ 
				\hline
  BTC & 0.0028 & 0.0454 & 0.0452 & 2.8227 & -0.1892 & 0.2276 \\ 
  ETH & 0.0019 & 0.0594 & 0.1501 & 2.0517 & -0.2228 & 0.2602 \\ 
  XRP & 0.0028 & 0.0767 & 1.6053 & 10.3886 & -0.3671 & 0.6183 \\ 
GLD & 0.0002 & 0.0062 & 0.1681 & 1.0159 & -0.0172 & 0.0254 \\ 
  SP500 & 0.0004 & 0.0086 & -0.5997 & 5.1430 & -0.0418 & 0.0484 \\ 
	\hline
			\hline
			&&&&&&
						\href{https://github.com/QuantLet/UCC}{\includegraphics[keepaspectratio,width=0.4cm]{media/qletlogo_tr.png}}
		\end{tabular}
	\end{table}

\vspace{-0.5cm}

(github.com/QuantLet/UCC)\\

\vspace{-1.3cm}

	\begin{figure}[H]
		\centering
		\hfill
		\subfigure{\includegraphics[width=7.5cm]{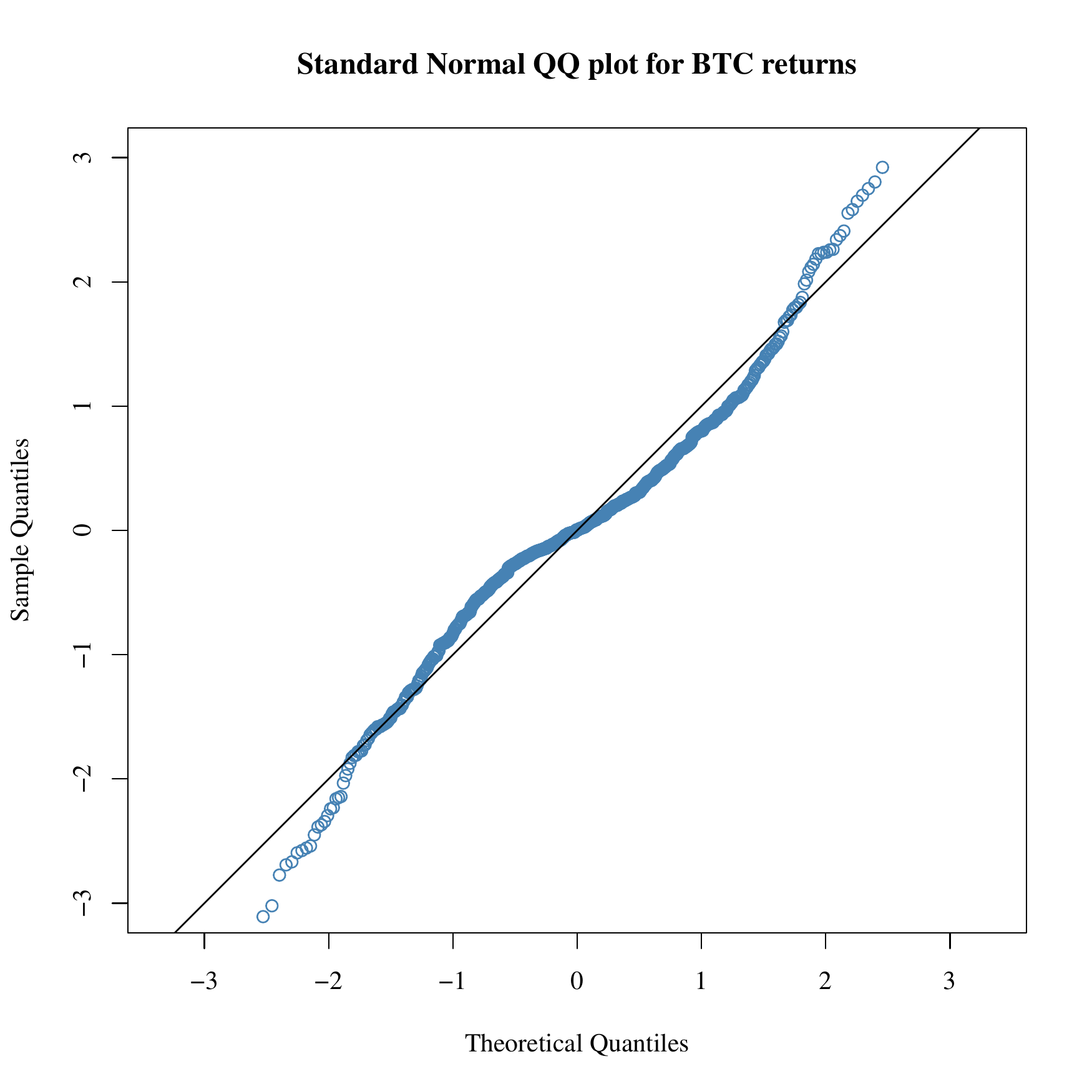}}
		\hfill
		\subfigure{\includegraphics[width=7.5cm]{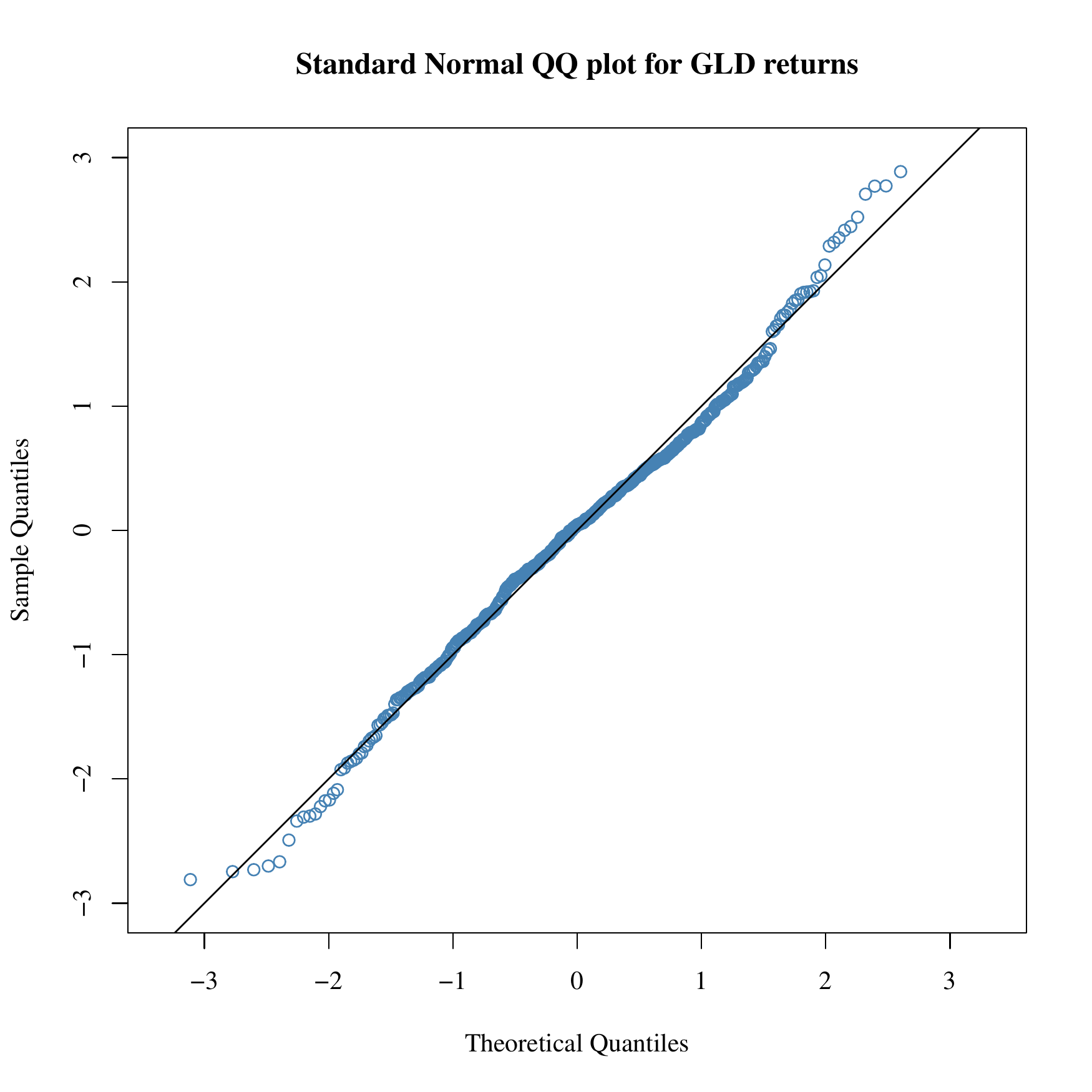}}
		\hfill
					\vspace{-0.5cm}

		\textcolor{white}{XXXXXXXXXXXXXXXXXXXXXXXXXXXXXXXXXXXXXXXXXXXXXLXX}\href{https://github.com/QuantLet/UCC}{\includegraphics[keepaspectratio,width=0.4cm]{media/qletlogo_tr.png}}
				\caption{Standard Normal QQ plots for BTC and GOLD, May 1, 2017 to Jun. 30, 2019. (github.com/QuantLet/UCC)}

	\end{figure}

	\vspace{-0.3cm}



	\begin{table}[H]
	\centering
	\caption{Johansen cointegration of gold and bitcoin, May 1, 2017 to Jun. 30, 2019.}
	\begin{tabular}{crrr}
		\hline
		\hline
		$H_0$             & Test statistic & 10 \% & 5 \%  \\ \hline
		$r$ $\le$ 1 & 3.11         & 7.52  & 9.24  \\
		$r$ = 0          & 18.22         & 13.75 & 15.67 \\ 
		\hline
		\hline
	\end{tabular}
\end{table}

\vspace{-0.5cm}
\textcolor{white}{XXX}Variable $r$ corresponds to the number of cointegration relations.  \href{https://github.com/QuantLet/UCC}{\includegraphics[keepaspectratio,width=0.4cm]{media/qletlogo_tr.png}}

(github.com/QuantLet/UCC)\\

	\vspace{0.5cm}

While cryptocurrencies have many similarities to gold (e.g., no central supply, no official price, and they can be mined), the correlation analysis suggests very little evidence of co-movement - at least over our limited sample. Not surprisingly, as shown in Table 4, we find no evidence that a cointegrating relationship exists (Johansen and Juselius, 1990; Johansen, 1991; Dwyer, 2015). Our findings are consistent with the findings of Klein et al. (2018), see subsection 4.6.

	\begin{table}[H]
	\centering
	\caption{Onatski and Wang cointegration of gold and bitcoin, May 1, 2017 to Jun. 30, 2019.}
	\begin{tabular}{crrr}
		\hline
		\hline
		$H_0$             & Test statistic & 10 \% & 5 \%  \\ \hline
		$r$ $\le$ 1 & 0.52         & 6.50  & 8.18  \\
		$r$ = 0          & 8.12         & 12.91 & 14.90 \\ 
		\hline
		\hline
	\end{tabular}
\end{table}

\vspace{-0.5cm}
\textcolor{white}{XXX}Variable $r$ corresponds to the number of cointegration relations. \href{https://github.com/QuantLet/CRIX_Cointegration}{\includegraphics[keepaspectratio,width=0.4cm]{media/qletlogo_tr.png}}

	\vspace{0.5cm}

\vspace{-0.5cm}

(github.com/QuantLet/UCC)\\

By testing our data with a new method suitable for high-dimensional nonstationary time series, as researched by Onatski and Chen (2018), we can underline the previous finding of the non-existence of a cointegrating relationship as shown in Table 5.

	
	\section{Potential Research Areas}
	\label{sec4}
	
Cryptocurrencies, as a new type of asset, offer many research opportunities for financial econometrics. For example, research on the dynamics of cryptocurrency trading, pricing, and volatility forecasting is advancing at a rapid pace (Briere et al., 2013; Gronwald, 2014; Cheung et al., 2015; Fry and Cheah, 2016; Chan et al., 2017; Kim et al., 2019). We will focus on the areas of network design, sentiment, and valuation; monetary systems and financial development; institutions; adoption, price discovery and high-frequency data; index construction; portfolio diversification; bubbles; alternative methods to raise capital; and the role of energy in consensus mechanisms.

	\subsection{Network design, sentiment, and valuation}
	\label{sec4.1}

The acceptance of this new technology has risen rapidly and activity in cryptocurrency trading has led to the establishment of more than 200 highly fragmented, mostly unregulat\-ed cryptocurrency exchanges, which act more like broker-dealers than traditional exchang\-es (Hansen, 2018). There is considerable ``off chain" trading. This might be intrabroker trading matching or even $dark$ $pool$ trading which might lead to price jumps at exchanges (Sharma, 2018). More types of cryptocurrency are being traded in parallel, on different exchanges, with different prices (see subsections 3.1 and 3.2.). These parallel information sources yield dynamic high-dimensional interdependencies.\\

Robinson et al. (2019) introduce a ``cross chain" technique which allows transactions to be executed and their respective value to be validated across sidechains.  They outline a programming model of a swap contract for exchanging value between sidechains, and discuss how this technology can be readily applied to many blockchain systems to provide cross-blockchain transactions.\\

A major problem for blockchain applications is the respective networks' $scalability$. Bu et al. (2019) research a distributed ledger system run by targeting algorithms that ensure a high throughput for the transactions generated in Internet-of-Things (IoT) systems. Transactions are continuously appended to an acyclic structure called tangle and each new transaction selects as parents two existing transactions (called tips) that it approves. This new metamorphic algorithm for tip selection by approving left behind tips, and improving confidence within the main tangle offers the best guaranties of both constructions called \href{https://www.iota.org/}{IOTA} (iota.org) and its proposed improved version G-IOTA.\\

A number of papers study the economic incentives of current consensus methods (Huberman et al., 2017). Biais et al. (2018) model the proof-of-work blockchain protocol as a stochastic game and analyse the equilibrium strategies of rational, strategic miners. They show how forks can be generated by information delays and software upgrades, and identify negative externalities. Easley et al. (2018) investigate the role that transaction fees play in the evolution of bitcoin from a mining-based structure to a market-based ecology. They develop a game-theoretic model to explain the factors leading to the emergence of transactions fees, as well as to explain the strategic behavior of miners and users. They highlight the role of mining rewards and trading volume, and examine how microstructure features such as exogenous structural constraints influence the dynamics and stability of the bitcoin blockchain. Cong et al. (2018a) develop a theory of mining pools that highlights risk sharing as a natural centralizing force.\\

Bhambhwani et al. (2019) research if cryptocurrencies have an intrinsic value related to the networks' computing power and network adoption. Their hypothesis is motivated by the fact that miners expend real resources to generate the computing power required to secure and operate the blockchain. An optimally performing blockchain serves as a medium for transactions and attracts users, developers, and intermediaries, thereby leading to an increase in the cryptocurrency’s network size. They find, that there is a positive and statistically
significant relationship among price, computing power, and network size (adoption levels respectively), which can be used
to construct asset pricing factors.\\

Ong et al. (2015) use social media data and find four key variables related to the market capitalization of a cryptocurrency: 1) merged pull requests on GitHub, 2) number of merges, 3) number of active accounts, and 4) number of total comments. The biggest cryptocurrencies by market capitalization experience the most activity, which is to be expected. However, the collection of this information is unique to cryptocurrencies. In the equity market, for example, similar information can be gleaned from third-party sources, like analyst reports and recommendations, and perhaps news flow and conference calls. The different sources of information available for cryptocurrencies presents new opportunities.\\
	
Research on trading patterns, herding effects, and economic decision making has started on sentiment construction/projection and cryptocurrency-specific lexica. Natural language processing techniques in combination with other machine learning techniques allow researchers to build sentiment measures. Cretarola and Figa-Talamanca (2017) propose a confidence-based model for asset and derivative prices in the bitcoin market with prices influenced by measures linked to the confidence in the underlying technology. Aste (2018) studies the dependency and causal structure of the current cryptocurrency market and investigates the collective movements of both prices and social sentiment related to almost 2,000 cryptocurrencies traded during the first six months of 2018. His results uncover a complex structure of interrelations, in which prices and sentiment influence each other across different currencies both instantaneously and with lead–lag relations.\\

Nasekin and Chen (2019) study investor sentiment on cryptocurrencies using a \newline cryptocurrency-specific lexicon proposed in Chen et al. (2018b) and statistical learning methods. Accounting for context-specific information and word similarity by learning word embeddings, they apply natural language processing methods for sentence-level classification and sentiment index construction. They argue that the constructed sentiment indices are value-relevant in terms of its return and volatility predictability for crypto\-currency market indices, see subsection 4.5. Pagnotta and Buraschi (2018) also address the valuation of cryptocurrencies, and characterize the demand for bitcoins by the available hashrate and show that the equilibrium price is obtained by solving a fixed-point problem. They find, that ``price/hashrate-spirals" amplify the demand and supply shocks.\\

 Schilling and Uhlig (2018) analyze the coexistence and the competition between the USD and bitcoin. They analyze bitcoin price evolution and interaction between the bitcoin price and monetary policy which	targets the USD, and obtain a fundamental pricing equation, which in its simplest form implies that bitcoin prices form a martingale.

	\subsection{Monetary systems and financial development}
	\label{sec4.2}

Blockchain-based monetary systems hold the potential to impact the macroeconomy, as the new payment systems challenge the traditional roles that banks have always played.
Cryptocurrencies may be viable competition for fiat currencies during periods when a central bank is perceived as weak or untrustworthy. However, the technology behind cryptocurrencies has the potential to improve a central banks' operations and can serve as a platform to launch their own cryptocurrencies (Raskin and Yermack, 2016). The petromondea (petro) issued by the government of Venezuela is an early example of these so-called central bank digital currencies (CBDC) (Keister and Sanches, 2018).\\

On June 18, 2019 Facebook announced to release a cryptocurrency on it's own in 2020 coined ``Libra".  Central authorities were fast to criticize this step and governments, including the United States, France, the United Kingdom, Germany, and Japan, expressed their resentment and scepticism. However, the propagation of systemic banking crises fostered by $too$-$big$-$to$-$fail$ financial institutions’ neverending propensity to take greater risks was a compelling reason behind the birth of cryptocurrencies over a decade ago and the design of Libra with its governance network of 28 high-profile firms already being in the project, looks similar to the Board of Governors of the Federal Reserve System comprising the twelve Federal Reserve Banks in the United States - with the difference of being in the hand of private actors and not a governement (Taskinsoy, 2019). \\

 Almosova (2018) studies the economics of blockchain currency systems and the respec\-tive value competition by applying a matching function of money demand to the operation of a blockchain to observe a monetary equilibrium. With cryptocurrencies already substitu\-ting for fiat money, Hendry and Zhu (2017) model the co-existence of different types of transactions and show that monetary authorities’ coordination capabilities are being restricted by the use of nonregulated cryptocurrencies.\\

The Catalini et. al (2019) paper on a market design for a blockchain-based financial system provides an extended abstract on a theory of long-run equilibrium in blockchain-based financial systems. Their theory elucidates the key market design features that separate proof-of-work and proof-of-stake approaches in the long run and when each design might each be appropriate (see subsection 2.4) and conclude, that with weak relational contracts or substantial concerns about outside interference, proof-of-work designs may be preferable. With regions that have local institutions that are reliable enough to make delegation feasible, proof-of-stake designs can lead to efficiency gains and improvements in governanc.\\
 	
 Cryptocurrencies may also increase financial inclusion and fuel economic activity in emerging markets, especially in sub-Saharan Africa, where only 34\% of adults had a bank account in 2014 (Blockchain Africa Conference 2018, blockchainafrica.co; GlobalFindex, 2017; World Bank, 2017). The possibility of using this technology for inclusion of the unbanked could allow billions to join the modern world of internet commerce and spur the creation of new businesses.

	
	\subsection{Institutions}
	\label{sec4.3}

There are hundred of cryptocurrency exchanges around the world. Some of the best-known are:  \href{https://www.binance.com/en}{Binance}, \href{https://www.bitfinex.com/}{Bitfinex} (bitfinex.com), \href{https://www.kraken.com}{Kraken} (kraken.com), \href{https://www.bitstamp.net/}{Bitstamp} (bitstamp.net), \href{https://www.coinbase.com/}{Coinbase} (coinbase.com),  \href{https://bitflyer.com/en-eu/}{Bitflyer} (bitflyer.com), \href{https://gemini24.zendesk.com}{Gemini} (gemini24.zendesk.com), \href{https://www.itbit.com/}{itBit} (itbit.com), \href{https://international.bittrex.com/}{Bittrex} (international.bittrex.com) and \href{https://poloniex.com/}{Poloniex} (poloniex.com). Each of these exchanges has its own specific traits. Kraken claims to be the largest bitcoin exchange in EUR volume and liquidity as well as being a partner in the first cryptocurrency bank, collaborating with the German \href{https://www.bafin.de/EN/Homepage/homepage_node.html}{BaFin} (bafin.de) -regulated bank Fidor. Shapeshift, in contrast, is an exchange that allows trades without signing up for an account. Gemini, being a fully US-regulated and licensed bitcoin and ethereum exchange, met its capital requirements by placing all USD deposits at a FDIC-insured bank.\\

 \href{https://coinmarketcap.com/exchanges/volume/24-hour/}{Coinmarketcap} (coinmarketcap.com/exchanges/volume/24-hour/) lists 218 exchanges, but even the measurement of trading volume is controversial. A recent filing to the SEC (2019) argues that 95\% of the trading volume in bitcoin is fake. The research identifies 10 exchanges with actual volume (out of 81), Binance, Bitfinex, Kraken, Bitstamp, Coinbase, Bitflyer, Gemini, itBit, Bittrex and Poloniex.\\

Due to the large number of exchanges with an ever increasing number of crypto\-currencies, price discrepancies due to market inefficiencies inherently exist. The low level of regulation and sentiment driven prices make pricing discrepancies larger than in other financial markets, such as fiat currency exchanges and stock exchanges. However, when one goes outside the ten exchanges with credible volume, some of the price discrepancies may not be real. Bistarelli et al. (2019) show via a theoretical model and via an empirical strategy, that arbitrage opportunities are possible by trading on different exchanges (Cretarola et al., 2017). Their approach is complementary to other theoretical studies on bitcoin arbitrage such as Barker (2017) or Pieters and Vivanco (2015), where the researchers study triangular arbitrage with bitcoin, i.e., buying bitcoin in USD and selling them in RMB.\\

Makarov and Schoar (2018) observe large recurrent arbitrage opportunities in crypto\-currency prices relative to fiat currencies across exchanges. These opportunities often persist for several days or weeks, and the price dispersions exist even in the face of significant trading volumes on many of the exchanges. Makarov and Schoar find that spreads are much smaller when cryptocurrencies are traded against each other, suggesting that cross-border controls on fiat currencies play an important role in creating the arbitrage opportunities. By constructing a common component and an idiosyncratic component, they conclude that the order flow plays an important role in explaining the spreads between exchanges. Further research in regards to arbitrage in bitcoin markets is done by Krueckeberg and Scholz (2018).\\

Bistarelli et al. (2018) show that cryptocurrency arbitrage strategies are profitable because the exchanges have different prices in the short run. Indeed, the fragmentation of exchanges is ideal for high-frequency trading bots. Bloomberg (2017) reports that Chinese high-frequency traders have used algorithms to identify mispricings and arbitrage opportunities across numerous exchanges in China. However, later in 2017, China banned all cryptocurrency exchanges.\\

Hautsch et al. (2018) note that consensus protocols confront traders with random waiting times until the transfer of ownership is accomplished. This settlement process exposes arbitrageurs to price risk and imposes limits to arbitrage. They derive theoretical arbitrage boundaries under general assumptions and show, by using high-frequency bitcoin data, that these increase with expected latency, latency uncertainty, spot volatility, and risk aversion. They conclude, that settlement through decentralized systems induces non-trivial frictions affecting market efficiency and price formation.

	\subsection{Adoption, price discovery and high-frequency data}
	\label{sec4.4}
	
	Cong et al. (2018b) provide the first fundamentals-based dynamic pricing model of cryptocurrencies and platform tokens, taking into consideration the user-base externality and endogenous user	adoption. Because the expectation of token price appreciation induces more agents to join the platform, tokens capitalize future user adoption, generally enhancing welfare and reducing user-base volatility (Sockin and Xiong, 2018). Catalini and Gans (2019) show that entrepreneurs have an incentive to use subsequent product pricing choices to ensure that crypto tokens issued to fund start-up costs --- a subject we discuss in subsection 4.8 --- retain their value even when they do not confer the typical rights associated with equity.\\
	
		Athey et al. (2016) develop a theoretical framework for bitcoin adoption and bitcoin pricing. Their paper relates to a variety of broad themes in the study of information technology adoption and usage. They conclude, that bitcoin presents a unique opportunity to observe both the adoption and micro-level user-to-user transaction and interaction data in the context of a new information technology product, and in an environment where the usage data is publicly available.\\
	
	Ghysels and Nguyen (2018) examine price discovery and liquidity provision in the secondary market for bitcoin and find that order informativeness generally increases with order aggressiveness, but that this pattern reverses in the outer layers of the book. Aggressive orders are more attractive to informed agents in a volatile market as reflected by the increased information content of such orders. They also find that market liquidity appears to migrate outward in response to the information asymmetry.\\

		 Griffin and Shams (2018) investigate whether tether --- see subsection 2.5 --- influences bitcoin and other cryptocurrency prices, and whether the growth of a pegged crypto\-currency is primarily driven by investor demand, or is supplied to investors as a scheme to profit from pushing cryptocurrency prices up. Their findings provide support for the view that price manipulation may be behind substantial distortive effects in cryptocurrencies.\\
	
	Wildi and Bundi (2018), analyze momentum trading strategies, and claim that bitcoin markets have become much more efficient markets. The impact of high-frequency trading combined with 24/7 trading opportunities has yet to be researched. It remains to be seen if an increase in liquidity will reduce the heretofore observed harsh swings in cryptocurrency prices.\\

	The launch of bitcoin futures on both the CME as well as the CBOE (XBT) provides opportunities to study price discovery (Karkkainen, 2018).
	Bitcoin futures trading gives many institutional investors the ability to invest in bitcoin and also allows to settle contracts in fiat money, potentially boosting liquidity. Currently the bitcoin futures volume is approximately the same as the largest exchange, Binance. Almost all of the volume is in the CME contract. The CBOE has announces they will no longer offer a bitcoin futures contract.\\
	
	Scaillet et al. (2018) identify high-frequency jump components in the bitcoin market and link them to new information arrival over time. Guo et al. (2018) perform a spectral clustering analysis of dynamic return-based network structures with coin attributions. This latent group structure in the cryptocurrency market leads them to conclude that comovements are influenced by the type of algorithm used. Makarov and Schoar (2018) study price deviations across cryptocurrency exchanges and interpret the deviations as the result of a balance between idiosyncratic sentiments of noise traders and the efforts of arbitrageurs to equilibrate prices across exchanges.

	\subsection{Index construction}
	\label{sec4.5}

Index construction poses unique challenges when it comes to cryptocurrencies. Traditional indices, such as the S\&P 500 or Russell 3000, gather data from stocks that are traded over particular time intervals in a small numbers of venues. Cryptocurrencies are traded 24/7 on hundreds of venues, like \href{https://www.worldcoinindex.com/}{WorldCoinIndex} (worldcoinindex.com),  \href{https://coinmarketcap.com/}{CoinMarketCap}, \href{https://www.cryptocompare.com/}{CryptoCompare}, \href{https://cci30.com/}{CryptoCurrencyIndex30} (cci30.com), or the  \href{https://www.cmegroup.com/trading/cryptocurrency-indices.html}{CME CF Cryptocurrency Indices} (cmegroup.com/trading/cryptocurrency-indices.html). Preliminary research on index construction is made, for example, by Trimborn and H\"ardle (2018), Chen et al. (2018a) and Kim et al. (2019).\\

While there are many exchanges, liquidity widely varies. Hence, the first challenge is what price should be used for individual cryptocurrencies. Even the original two bitcoin futures contracts (CME and CBOE) use different data sources for the price of bitcoin. Indeed, bitcoin is the most liquid cryptocurrency and there is no agreement on the ``spot" price. The difficulty in establishing a price and the possibility of price manipulation on certain exchanges has lead the SEC to block the creation of cryptocurrency ETFs.\\

Kim et al. (2019) have set the goal of capturing the expectations on the cryptocurrency market (represented by \href{https://thecrix.de/}{CRIX}) through the construction of an implied volatility proxy in absence of the derivatives for the majority of cryptocurrencies. The ``fear index" VIX of the United States stock market was selected as a guidance. Analysis of the relationships between VIX and
volatility of the underlying assets provide an insight for the selection of a respective proxy. The established VCRIX index provides a daily forecast for the mean annualized volatility of the next 30 days.\\

There are other issues that provide a challenge in index construction such as forking. When are forked cryptocurrencies added to the index? If the index focuses on large capitalized cryptocurrencies, should a smaller capitalized fork be included? Forks are a new concept that poses a challenge to financial engineering. The forking problem is also a challenge for the single currency futures contracts.

	\subsection{Portfolio diversification}
	\label{sec4.6}

	For millenia, gold has been an accepted store and measure of value, offering very long-term stability and security in the financial marketplace (Erb and Harvey, 2013). Bitcoin and gold are similar from both a psychological perspective and, especially, as a resource. Neither can be created arbitrarily: each must be mined and each has a finite supply (at least on planet Earth).	That said, gold has fundamental value when used for jewelry and art as well as electronic or medical components. The limited supply of ``digital gold", combined with the market’s current acceptance of it, suggests that bitcoin and other cryptocurrencies may be able to serve a similar role as gold. Klein et al. (2018) show that the volatility dynamics of cryptocurrencies do share some similarities with those of gold and silver.\\
	
	Gkillas and Longin (2018) argue that bitcoin is the new digital gold and they investigate the potential benefits of bitcoin during extremely volatile market periods. They find that the correlation of extreme returns between bitcoin and US and European equity markets increases during stock market drawdowns and decreases during stock market booms. Their conclusion is that bitcoin can play an important role in asset management and provide similar results as those of gold. Furthermore, Gkillas and Longin find a low extreme correlation between bitcoin and gold, implying that the assets can be used together in turbulent times. That said, we suggest caution in interpreting these results given the very limited data.\\

Petukhina et al. (2018) find that due to the volatility structure of cryptocurrencies, the application of traditional risk-based portfolios --- such as equal-risk contribution, minimum-variance and minimum-CVaR portfolios --- does not boost the performance of investments significantly. Liu et al. (2019a) examine common risk factors in crypto\-currencies, and capture the cross-sectional expected cryptocurrency returns. By con\-sidering a comprehensive list of price- and market-related factors in the stock market, they construct cryptocurrency counterparts. Their cryptocurrency factors claim to be successful long-short strategies that generate sizable and statistically significant excess returns. The paper thus establishes a set of
stylized facts on the cross-section of crypto\-currencies that can be used to assess and develop theoretical models.

	\subsection{Bubbles}
	\label{sec4.7}
	
	Chaim and Laurini (2019) analyze daily returns of bitcoin between January 2015 and March 2018 to empirically investigate the price bubble hypothesis. Bitcoin returns have characteristics one would expect of a bubble: it is very volatile, exhibits large kurtosis, and negative skewness (Camerer, 1989). By following previous research, they conclude that bitcoin-USD prices being a bubble is plausible, but the evidence is inconclusive.\\
	
	 In contrast, Henry and Irrera (2017) argue that cryptocurrencies exhibit bubble-like behavior. Recent research by Hafner (2018), contained in this special issue, extends traditional bubble tests to the case of time-varying volatility. Dong et al. (2018) investigate the positive and negative outcomes of a  cryptocurrency model as risky and costly bubbles in an infinite-horizon production economy with incomplete markets that has the following framework for bitcoin: 1) enormous volatility, 2) price dynamics are significantly sensitive to both investor sentiment and policy stances, and 3) the market exhibits diverse cyclical features for US and China. Their quantitative results, however, rely heavily on the severity of the market distortion, i.e. the intervention in the given market by a governing body, which, in
	 turn, determines the size of the bitcoin bubbles.\\

Shu and Zhu (2019) employ the log-periodic power law singularity (LPPLS) confidence indicator as a diagnostic tool for identifying bubbles using the daily data on bitcoin price. The LPPLS confidence indicator fails to provide effective warnings for detecting the bubbles when the bitcoin price suffers from a large fluctuation in a short time, especially for positive bubbles. In order to diagnose the existence of bubbles and accurately predict the bubble crashes in the cryptocurrency market, their research proposes an adaptive multilevel time series detection methodology based on the LPPLS model and high frequency data, which effectively detects bubbles and accurately forecasts bubble burts. On a day to week scale, the LPPLS confidence indicator has a stable performance in terms of effectively monitoring the bubble status on a longer time scale - on a week to month scale. Their adaptive multilevel time series detection methodology claims to provide real-time detection of bubbles and advanced forecast of crashes to warn of the imminent risk.

	\subsection{Alternative methods to raise capital}
	\label{sec4.8}
	
The year 2017 brought a surge in initial coin offerings (ICOs), similar to initial public offerings (SEC-approved stock offerings). ICO's are a potentially new financing channel for entrepreneurs (Cong and He, 2017). The space has also generated a lot of attention because some investors are buying into ICOs without fully understanding the technology as well as some companies are offering an ICO without an economically meaningful use case for the cryptocurrency (Ernst \& Young, 2017; Amsden and Schweitzer, 2018).\\

Indeed, cryptocurrencies hold the potential to significantly reduce cost, complexity, and simultaneously increase the speed of trading and settlement processes in a secure manner. Cryptocurrencies are tokens, but other assets such as shares of a company can similarly be tokenized and traded.\\

 In summary, 329 ICOs out of 2027 ICOs listed on \href{http://www.tokendata.io}{tokendata.io} have failed (16.23\%). Extensive research is maintained regarding ICOs (Santo et al., 2016; Bajpai, 2017; SEC, 2017a, 2017b; Adhami et al., 2018; Momtaz, 2018; Kostovetsky, 2018; Guegan and Henot, 2018; Howell et al., 2018; Liu et al., 2019b). In the sample used by Bourveau et al. (2018) approximately 85\% of ICOs are successful.\\ 
 
 Basic alternatives to traditional banking services that have a cryptocurrency backbone are being researched as well. Panwar et al. (2019) research a blockchain-based credit network where credit transfer between a sender-receiver pair happens on demand. Dis\-tributed credit networks (DCNs) are distributed systems of trust between users, where a user extends financial credit, or guarantees assets to other users whom it deems credit worthy, with the extended credit proportionate to the amount of trust that exists between the users --- essentially peer-to-peer lending networks, where users extend credit, borrow money and	commodities from each other directly, while minimizing the role of banks, clearing-houses, or bourses. They present preliminary experiments and scalability analyses based on their proposed DCN framework.

	\subsection{The role of energy in consensus mechanisms}
	\label{sec4.9}
	
	As we have detailed, there are many different consensus mechanisms. Bitcoin uses a particularly energy-intensive method, which raises environmental concerns, especially with the prevalence of bitcoin mining dependent on coal-fired power plants in China (Hileman and Rauchs, 2017). Cong et al. (2018a) show that mining pools, as a financial innovation, significantly exacerbate energy consumption for proof-of-work-based block\-chains in their research output regarding decentralized mining in centralized pools. As of April 2018, aggregate energy devoted to bitcoin mining alone exceeded 60 TWh, roughly the annual energy consumed by Switzerland as a country (Lee, 2018). Mishra et al. (2018) investigate how the mining protocol of bitcoin impacts the computing capacity needs of miners and demonstrate, that the mining algorithm as well as the transaction volume increase computing resource needs, which in turn raises the energy consumption. Eventually they argue resource requirements both from a computing hardware and energy consumption needs that the future growth of the bitcoin network and the use of bitcoin as a currency could be questionable.\\

As the annual electricity consumption for cryptocurrency mining is growing yearly. Total carbon production from mining now likely exceeds that generated by the entire nation of Portugal. Corbet et al. (2019) investigate how Bitcoin's price volatility and the underlying dynamics of cryptocurrency's mining characteristics affect the energy markets, utilities companies, and green ETFs. The results claim that continued cryptocurrency energy-usage impacts the performance of energy sector, which emphasises the importance of further assessment of environmental impacts of cryptocurrency growth. Blockchain technology offers a number of innovative environment-related research opportunities\linebreak (Hayes, 2017; Pop et al., 2018).

	\section{Closing Remarks}
	\label{sec5}
	
Cryptocurrencies are an intriguing financial innovation and offer many possible research avenues. As with many new technologies, considerable confusion exists about both the underlying concept of cryptocurrencies and the approaches for valuing them.\\

Our first goal in this paper is to provide a high level understanding of the blockchain technology behind the cryptocurrencies. Second, we want to emphasize that there are many different classes of cryptocurrencies --- too often cryptocurrency is summarized as bitcoin. Cryptocurrencies vary, however, and can be tokens representing shares of traditional assets, provide direct utility such as computational power, and even represent a fiat currency. \\

Finally, we would like to emphasize the large number of research avenues available in the cryptocurrency space. In 2018, we witnessed the bursting of a bubble in the most liquid cryptocurrencies, but the research opportunities go well beyond bubbles. There is much to do in this new field of finance and economics.


	\section{References}
	\label{sec6}
	
	\begin{footnotesize}
		
		\raggedright
		
	Abadi, J., and M. Brunnermeier. 2018. ``Blockchain Economics". Princeton University. Retrieved on the 24.01.2019 from \url{https://scholar.princeton.edu/sites/default/files/markus/files/blockchain_paper_v3g.pdf}.\vspace{0.2cm}
	
		Abraham, I., D. Malkhi, K. Nayak, L. Ren, and A. Spiegelman. 2016. ``Solidus: An incentive-compatible cryptocurrency based on permissionless Byzantine consensus." Retrieved on the 24.11.2018 from \url{https://arxiv.org/abs/1612.02916}.\vspace{0.2cm}
		
		Adhami, S., G. Giudici, and S. Martinazzi 2018. ``Why Do Businesses Go Crypto? An
Empirical Analysis of Initial Coin Offerings." Journal of Economics and Business. Forthcoming. \vspace{0.2cm}
		
		Almosova, A. 2018. ``A Monetary Model of Blockchain." IRTG 1792 DP 2018-008. Retrieved on the 12.02.2018 from \url{https://www.wiwi.hu-berlin.de/de/forschung/irtg/results/discussion-papers/discussion-papers-2017-1/irtg1792dp2018-008.pdf}.\vspace{0.2cm}
				
		Ametrano, F. 2016. ``Hayek money: The cryptocurrency price stability solution." Retrieved on the 24.11.2018 from \url{http://dx.doi.org/10.2139/ssrn.2425270}.\vspace{0.2cm}
		
		Amsden, R., and D. Schweizer 2018. ``Are Blockchain Crowdsales the New 'Gold Rush'?
Success Determinants of Initial Coin Offerings." Retrieved on the 12.10.2018 from \url{https://papers.ssrn.com/sol3/papers.cfm?abstract_id=3163849}\vspace{0.2cm}

Athey, S., Parashkevov, I., Sarukkai, V., and J. Xia. 2016. ``Bitcoin Pricing, Adoption, and Usage: Theory and Evidence". Stanford University. Retrieved on the 11.03.2019 from \url{https://siepr.stanford.edu/sites/default/files/publications/17-033_1.pdf}.\vspace{0.2cm}
		
		Aranda, D., and R. Zagone. 2015. ``The ‘Ripple’ Effect: Why an Open Payments Infrastructure Matters." Consultative Group to Assist the Poor. Retrieved on the 26.11.2018 from \url{http://www.cgap.org/blog/\%E2\%80\%98ripple\%E2\%80\%99-effect-why-open-payments-infrastructure-matters}.\vspace{0.2cm}
		
		Arnold, M. 2018. ``Ripple and Swift slug it out over cross-border payments." Financial Times. Retrieved on the 25.08.2018 from \url{https://www.ft.com/content/631af8cc-47cc-11e8-8c77-ff51caedcde6}.\vspace{0.2cm}

Aste, T. 2018. ``Cryptocurrency market structure: connecting emotions and economics. Special Issue of Digital Finance on Cryptocurrencies." Digital Finance. Smart Data Analytics, Investment Innovation, and Financial Technology. ISSN 2524-6186.\vspace{0.2cm}

Babich, V., and G. Hilary. 2018a. ``Blockchain and Other Distributed Ledger Technologies in Operations". SSRN. Retrieved on the 15.01.2019 from \url{https://papers.ssrn.com/sol3/papers.cfm?abstract_id=3232977}.\vspace{0.2cm}

Babich, V., and G. Hilary. 2018b. ``Distributed Ledgers and Operations: What Operations
Management Researchers Should Know about Blockchain Technology." Georgetown McDonough School of Business Research Paper No. 3131250.\vspace{0.2cm}

		Back, A. 2002. ``Hashcash - A Denial of Service Counter-Measure." Retrieved on the 24.12.2018 from \url{http://www.hashcash.org/papers/hashcash.pdf}.\vspace{0.2cm}
		
		Bajpai, P. 2008. Bitcoin: A peer-to-peer electronic cash system. Retrieved on the 24.11.2018 from \url{http://www.bitcoin.org/bitcoin.pdf}.\vspace{0.2cm}
		
		Barker, J. 2017. ``Triangular arbitrage with Bitcoin." Texas A\&M University. Retrieved on the 24.02.2019 from \url{https://oaktrust.library.tamu.edu/bitstream/handle/1969.1/164589/BARKER-DOCUMENT-2017.pdf?sequence=1&isAllowed=y}.\vspace{0.2cm}
		
		Bartos, J. 2015. ``Does bitcoin Follow the Hypothesis of Efficient Market?" International Journal of Economic Sciences, IV, 2, 10–23.\vspace{0.2cm}
		
		Bech, M., and R. Garratt. 2017. ``Central bank cryptocurrencies". BIS Quarterly Review. Volume 55, September 2017.\vspace{0.2cm}
		
		Benedetti, H., and L. Kostovetsky. 2018. ``Digital Tulips? Returns to Investors in Initial Coin Offerings". SSRN. Retrieved on the 16.02.2019 from 
		\url{https://papers.ssrn.com/sol3/papers.cfm?abstract_id=3182169}.\vspace{0.2cm}
		
		Biais,	B., Bisière, Ch., Bouvard, M.,	and	C. Casamatta. 2018. ``The	blockchain	folk	theorem". SSRN. Retrieved on the 08.01.2019 from 
		\url{https://papers.ssrn.com/sol3/papers.cfm?abstract_id=3108601}.\vspace{0.2cm}
		
		Bistarelli, S., A. Cretarola, G. Figa-Talamanca, and M. Patacca. 2018. ``Model-based arbitrage in multi-exchange models for bitcoin price dynamics." Special Issue of Digital Finance on Cryptocurrencies. Digital Finance. Smart Data Analytics, Investment Innovation, and Financial Technology. ISSN 2524-6186.\vspace{0.2cm}
		
		Bistarelli, S., A. Cretarola, G. Figà-Talamanca, I. Mercanti, and M. Patacca. 2019. ``Is Arbitrage Possible in the Bitcoin Market?" (Work-In-Progress Paper). In: M. Coppola, E. Carlini, D. D’Agostino, J. Altmann, and J. Bañares (eds) Economics of Grids, Clouds, Systems, and Services. GECON 2018. Lecture Notes in Computer Science, vol 11113. Springer, Cham.\vspace{0.2cm}
		
		Bloomberg. 2017. ``High-Speed Traders Are Taking Over bitcoin." Bloomberg. Retrieved on the 16.02.2018 from\\
		\url{https://www.bloomberg.com/news/articles/2017-01-16/high-speed-traders-are-taking-over-bitcoin-as-easy-money-beckons}.\vspace{0.2cm}
		
		B\"ohme, R., N. Christin, B. Edelman, and T. Moore. 2015. ``bitcoin: Economics, technology, and governance." Journal of Economic Perspectives, 29(2), pp.213-38.\vspace{0.2cm}
		
		Bordo, M., Levin, A. 2017. ``Central Bank Digital Currency and the Future of Monetary Policy". National Bureau of Economic Research. NBER Working Paper No. 23711.\vspace{0.2cm}
		
		Borke, L., and W. H\"ardle 2018. ``Q3-D3-LSA." In H\"ardle, W., Lu, H., Shen, X. (Eds.). Handbook of Big data Analytics. Springer-Verlag Berlin Heidelberg. ISBN 978-3-319-18284-1, DOI: 10.1007/978-3-319-18284-1.\vspace{0.2cm}
		
		Bourveau, T., De George, E., Ellahie, A., and D. Macciocchi. 2018. Initial Coin Offerings: Early Evidence on the Role of Disclosure in the Unregulated Crypto Market. SSRN. Retrieved on the 24.11.2018 from \url{https://papers.ssrn.com/sol3/papers.cfm?abstract_id=3193392}.\vspace{0.2cm}
		
		Briere, M., K. Oosterlinck, and A. Szafarz. 2013. ``Virtual Currency, Tangible Return: Portfolio Diversification with bitcoins." Retrieved on the 24.11.2018 from: \url{https://econpapers.repec.org/paper/solwpaper/2013_2f149159.htm}.\vspace{0.2cm}
		
Bu, G., Hana, W., and M. Potop-Butucaru. 2019. ``Metamorphic IOTA." Retrieved on the 18.06.2019 from \url{https://arxiv.org/pdf/1907.03628.pdf}.\vspace{0.2cm}

		Bundesbank. 2017. ``Monatsbericht: Bundesbank sieht Vorteile in Blockchain-Technologie." Deutsche Bundesbank. Retrieved on the 14.12.2018 from \url{https://www.bundesbank.de/Redaktion/DE/Themen/2017/2017_09_18_monatsbericht_dlt.html/}.\vspace{0.2cm}
		
		Buterin, V. 2014. ``On Stake. Ethereum Blog." Retrieved on the 24.11.2018 from \url{https://blog.ethereum.org/2014/07/05/stake/}.\vspace{0.2cm}

		 Bhambhwani, S., Delikouras, S. and G. Korniotis. 2019. ``Do Fundamentals Drive Cryptocurrency Prices?" SSRN. Retrieved on the 30.07.2019 from \url{https://ssrn.com/abstract=3342842}.\vspace{0.2cm}
		
		Camerer, C. 1989. ``Bubbles and fads in asset prices." Journal of Economic
		Surveys 3 (1): 3–41. DOI: https://doi.org/10.1111/j.1467-6419.1989.tb00056.x.\vspace{0.2cm}

		Catalini, Ch., and J. Gans. 2016. ``Some Simple Economics of the Blockchain." MIT Sloan Research Paper No. 5191-16. Retrieved on the 12.05.2018 from \url{http://dx.doi.org/10.2139/ssrn.2874598}.\vspace{0.2cm}

		Catalini, Ch., Jagadeesan, R., and S. Duke Kominers. 2019. `` Market Design for a Blockchain-Based Financial System." SSRN. Retrieved on the 10.07.2019 from \url{https://papers.ssrn.com/sol3/papers.cfm?abstract_id=3396834}.\vspace{0.2cm}
		
		Catalini, Ch., and J. Gans. 2019. ``Initial Coin Offerings and the Value of Crypto Tokens." MIT Sloan Research Paper No. 5347-18. Retrieved on the 11.03.2019 from \url{http://dx.doi.org/10.2139/ssrn.3137213}.\vspace{0.2cm}
		
		Chan, S., J. Chu, S. Nadarajah, and J. Osterrieder. 2017. ``A Statistical Analysis of Cryptocurrencies." Journal of Risk and Financial Management, 2017, 10(2), 12.\vspace{0.2cm}
		
		Chaim, P., and M. Laurini. 2019. ``Is Bitcoin a bubble?" Physica A: Statistical Mechanics and its Applications. Volume 517, 1 March 2019, Pages 222-232. DOI: https://doi.org/10.1016/j.physa.2018.11.031.\vspace{0.2cm}
		
		Chaum, D. 1983. ``Blind signatures for untraceable payments." Advances in Cryptology Proceedings of Crypto. 82 (3): 199–203. doi:10.1007/978-1-4757-0602-4\_18.\vspace{0.2cm}
		
		Cheung, A., E. Roca, and J. Su. 2015. ``Crypto-currency bubbles: an application of the Phillips–Shi–Yu (2013) methodology on Mt. Gox bitcoin prices." Applied Economics, 47(23), 2348-2358.\vspace{0.2cm}
		
		Chen, C. YH., W. Härdle, A. Hou, and W. Wang. 2018a. ``Pricing Cryptocurrency Options: The Case of CRIX and Bitcoin". Journal of Financial Econometrics. Contained in this special issue.\vspace{0.2cm}
		
		Chen, C. YH., R. Després, L. Guo, and T. Renault. 2018b. ``What makes cryptocurrencies special? Investor sentiment and price predictability in the absence of fundamental value." IRTG 1792 Discussion Paper. ISSN: 2568-5619. Forthcoming.\vspace{0.2cm}
		
Cretarola, A., G. Figa-Talamanca. 2017. ``A Confidence-Based Model for Asset and Derivative Prices in the bitcoin Market." Retrieved on the 24.01.2018 from \url{http://dx.doi.org/10.2139/ssrn.2908921}.\vspace{0.2cm}

Cretarola, A., Figà-Talamanca, G., and M. Patacca. 2017. ``A Sentiment-Based Model for the Bitcoin: Theory, Estimation and Option Pricing." SSRN. Retrieved on the 19.03.2019 from \url{https://papers.ssrn.com/sol3/papers.cfm?abstract_id=3042029}.\vspace{0.2cm}

Cong, L., and He, Z. 2017. ``Blockchain Disruption and Smart Contracts". SSRN. Retrieved on the 24.11.2018 from \url{https://papers.ssrn.com/sol3/papers.cfm?abstract_id=2985764}.\vspace{0.2cm}
		
		Cong, L., He, Z., and J. Li. 2018a. ``Decentralized Mining in Centralized Pools". SSRN. Retrieved on the 24.11.2018 from \url{https://papers.ssrn.com/sol3/papers.cfm?abstract_id=3143724}.\vspace{0.2cm}

		Cong, L., Li, Y., and N. Wang. 2018b. ``Tokenomics: Dynamic Adoption and Valuation
		". SSRN. Retrieved on the 24.02.2019 from \url{https://papers.ssrn.com/sol3/papers.cfm?abstract_id=3153860}.\vspace{0.2cm}

		Corbet, S., Lucey, B., and L. Yarovaya. 2019. ``The Financial Market Effects of Cryptocurrency Energy Usage". SSRN. Retrieved on the 24.06.2019 from \url{https://papers.ssrn.com/sol3/papers.cfm?abstract_id=3412194}.\vspace{0.2cm}

		Cotillard, M. 2015. ``bitcoin's Block Size Debate Tests Its Community Governance." Brave New Coin. 18 August. Retrieved on the 24.11.2018 from \url{https://bravenewcoin.com/news/bitcoins-block-size-debate-tests-its-community-governance/}.\vspace{0.2cm}
		
		Delmolino, K., M. Arnett, A. Kosba, A. Miller, and E. Shi. 2016. ``Step by step towards creating a safe smart contract: Lessons and insights from a cryptocurrency lab." Retrieved on the 24.11.2018 from \url{https://eprint.iacr.org/2015/460.pdf}.\vspace{0.2cm}
		
		Derose, C. 2015. ``Blockchain for Beginners - Behind the Ingenious Security Feature that Powers the Blockchain." American Banker. Retrieved on the 24.11.2018 from \url{https://www.americanbanker.com/opinion/behind-the-ingenious-security-feature-that-powers-the-blockchain}.\vspace{0.2cm}
		
		Dong, F., Z. Xu, and Y. Zhang. 2018. ``Bubbly bitcoin." SSRN. Retrieved on the 07.01.2019 from \url{https://papers.ssrn.com/sol3/papers.cfm?abstract_id=3290125}.\vspace{0.2cm}
		
		Dwork, C., and M. Naor. 1992. ``Pricing via Processing or Combatting Junk Mail." Annual International Cryptology Conference. CRYPTO 1992: Advances in Cryptology. CRYPTO 92, pp. 139-147.\vspace{0.2cm}
			
		Dwyer, G. 2015. ``The Johansen Tests for Cointegration." Retrieved on the 24.02.2018 from \url{http://www.jerrydwyer.com/pdf/Clemson/Cointegration.pdf}.\vspace{0.2cm}

Easley, D., O'Hara, M, and S. Basu. 2018. ``From Mining to Markets: The Evolution of Bitcoin Transaction Fees". Journal of Financial Economics (JFE). Forthcoming.\vspace{0.2cm}

Erb, C., and C. Harvey. 2013. ``The Golden Dilemma." Financial Analysts Journal, 2013:69:4, July-August, pp. 10-42.\vspace{0.2cm}

Ernst \& Young. 2017. ``EY Research: Initial Coin Offerings (ICOs)."
Retrieved on the 12.12.2018 from \url{https://www.ey.com/Publication/vwLUAssets/ey-research-initial-coin-offerings-icos/$File$/ey-research-initial-coin-offerings-icos.pdf}.\vspace{0.2cm}
		
		Franco, P. 2015. ``Understanding bitcoin." Chichester, West Sussex: John Wiley \& Sons Ltd.\vspace{0.2cm}
		
		Foley, S., Karlsen, J., and T. Putnins. 2018. ``Sex, Drugs, and Bitcoin: How Much Illegal Activity Is Financed Through Cryptocurrencies?". Review of Financial Studies, Forthcoming.\vspace{0.2cm}

		Fry, J., and E. Cheah. 2016. ``Negative bubbles and shocks in cryptocurrency markets." International Review of Financial Analysis, 47, 343-352.\vspace{0.2cm}

Ghysels, E., and G. Nguyen. 2018. ``Price Discovery of a Speculative Asset: Evidence from a bitcoin Exchange." SSRN. Retrieved on the 15.10.2018 from \url{https://papers.ssrn.com/sol3/papers.cfm?abstract_id=3258508}.\vspace{0.2cm}

Gkillas, K., and F. Longin. 2018. ``Is bitcoin the New Digital Gold? Evidence From Extreme Price Movements in Financial Markets." SSRN. Retrieved on the 15.10.2018 from \url{https://papers.ssrn.com/sol3/papers.cfm?abstract_id=3245571}.\vspace{0.2cm}

		GlobalFindex. 2017. ``The Global Findex Database 2017." The World Bank. Retrieved on the 05.06.2018 from \url{https://globalfindex.worldbank.org}.\vspace{0.2cm}

		Gronwald, M. 2014. ``The Economics of bitcoins - Market Characteristics and Price Jumps." CESifo Working Paper, No. 5121.\vspace{0.2cm}
		
		Griffin, J., and A. Shams. 2018. ``Is Bitcoin Really Un-Tethered?". SSRN. Retrieved on the 11.03.2019 from \url{https://papers.ssrn.com/sol3/papers.cfm?abstract_id=3195066}.\vspace{0.2cm}

		Guegan, D., and C. Henot 2018. A Probative Value for Authentication Use Case Blockchain. Special Issue of Digital Finance on Cryptocurrencies. Digital Finance. Smart Data Analytics, Investment Innovation, and Financial Technology. ISSN 2524-6186. Submitted.\vspace{0.2cm}

		Guo, L., and X. Li. 2017. ``Risk Analysis of Cryptocurrency as an Alternative Asset Class." In H\"ardle, W., Chen, C. YH., Overbeck, L. (Eds.). Applied Quantitative Finance. Third edition. Springer-Verlag Berlin Heidelberg. ISBN 978-3-662-54485-3, e-ISBN 978-3-662-54486-0, DOI:10.1007/978-3-662-54486-0, pp. 309-329.\vspace{0.2cm}
		
		Guo, L., Y. Tao, and W. H\"ardle. 2018. ``A Dynamic Network for Cryptocurrencies." Journal of the American Statistical Association. Submitted.\vspace{0.2cm}
		
		Haber, S., and W. Stornetta. 1991. ``How to Time-Stamp a Digital Document." Journal of Cryptology. Volume 3, Issue 2, pp. 99–111\vspace{0.2cm}
		
		Hafner, Ch. 2018. ``Testing for Bubbles in Cryptocurrencies with Time-Varying Volatility."  Journal of Financial Econometrics. Contained in this special issue.\vspace{0.2cm}		
		
		Hansen, S. 2018. ``Guide To Top Cryptocurrency Exchanges." Forbes. Retrieved on the 15.08.2018 from \url{https://www.forbes.com/sites/sarahhansen/2018/06/20/forbes-guide-to-cryptocurrency-exchanges/#503cf5ce2572}\vspace{0.2cm}
		
Harvey, C. 2014. ``Bitcoin Myths and Facts." Retrieved on the 26.11.2018 from \url{https://ssrn.com/abstract=2479670}\vspace{0.2cm}

Harvey, C. 2016. ``Cryptofinance." Retrieved on the 9.02.2018 from \url{https://papers.ssrn.com/sol3/papers.cfm?abstract_id=2438299}\vspace{0.2cm}	

Harvey, C. 2017a. ``Breaking Down bitcoin." Retrieved on the 26.05.2018 from \url{https://www.fuqua.duke.edu/duke-fuqua-insights/breaking-down-bitcoin-\%E2\%80\%93-professor-campbell-harvey-digital-currency\%E2\%80\%99s-prospects}\vspace{0.2cm}		

Harvey, C. 2017b. ``Blockchain 2.0." Global Risk Institute. Summit 2017. Retrieved on the 12.03.2018 from \url{http://globalriskinstitute.org/wp-content/uploads/2017/10/Campbell-Harvey-Blockchain-2.0.pdf}\vspace{0.2cm}

Hautsch, N., Scheuch, Ch., and S. Voigt. 2018. ``Limits to Arbitrage in Markets With Stochastic Settlement Latency." CFS Working Paper, No. 616, 2018.\vspace{0.2cm}

Hayes, A. 2017. ``Cryptocurrency Value Formation: An empirical study leading to a cost of production model for valuing Bitcoin." Telematics and Informatics. 34. 1308-1321. 10.1016/j.tele.2016.05.005.\vspace{0.2cm}
		
		Hendry, S., and Y. Zhu. 2017. ``A Framework for Analyzing Monetary Policy in an Economy
with Emoney." Bank of Canada. Retrieved on the 12.11.2018 from \url{https://editorialexpress.com/cgi-bin/conference/download.cgi?db_name=mmmspr2017&paper_id=104}\vspace{0.2cm}
		
				Henry, D. and A. Irrera. 2017. ``JPMorgan's Dimon says bitcoin 'is a fraud'." Retrieved on the 15.12.2018 from \url{https://www.reuters.com/article/legal-us-usa-banks-conference-jpmorgan/jpmorgans-dimon-says-bitcoin-is-a-fraud-idUSKCN1BN2PN}\vspace{0.2cm}

	Hileman, G., and M. Rauchs. 2017. ``Global Cryptocurrency Benchmarking Study". Retrieved on the 06.01.2018 from\\ \url{https://www.jbs.cam.ac.uk/fileadmin/user_upload/research/centres/alternative-finance/downloads/2017-global-cryptocurrency-benchmarking-study.pdf}.\vspace{0.2cm}
	
	Howell, S., Niessner, M., and D. Yermack. 2018. ``Initial Coin Offerings: Financing Growth with Cryptocurrency Token Sales". European Corporate Governance Institute (ECGI) - Finance Working Paper No. 564/2018.\vspace{0.2cm}
	
	Hu, A., Parlour, C., and U. Rajan. 2018. ``Cryptocurrencies: Stylized Facts on a New Investible Instrument". SSRN. Retrieved on the 06.03.2019 from  \url{https://papers.ssrn.com/sol3/papers.cfm?abstract_id=3182113}.\vspace{0.2cm}
	
	Huberman, G., Leshno, J., and C. Moallemi. 2017. ``Monopoly Without a Monopolist: An Economic Analysis of the Bitcoin Payment System". Bank of Finland Research Discussion Paper No. 27/2017.\vspace{0.2cm}
		
		Iwamura, M., Y. Kitamura, and T. Matsumoto. 2014. ``Is bitcoin the Only Cryptocurrency in the Town?" Economics of Cryptocurrency and Friedrich A. Hayek. Retrieved on the 24.11.2018 from \url{https://econpapers.repec.org/paper/hithituec/602.htm}.\vspace{0.2cm}
		
		Jakobsson, M., and A. Juels. 1999. ``Proofs of Work and Bread Pudding Protocols. Communications and Multimedia Security." Kluwer Academic Publishers: 258–272.\vspace{0.2cm}
		
		Johansen, S., and K. Juselius. 1990. ``Maxium likelihood estimation and
		inference on cointegration - with applications to the demand for money." Oxford Bulletin of economics and statistics. Vol. 52, Issue 2, pp. 169-210.\vspace{0.2cm}
		
		Johansen, S. 1991. ``Estimation and Hypothesis Testing of Cointegration Vectors in Gaussian Vector Autoregressive Models." Econometrica 59: 1551–1580.\vspace{0.2cm}
		
		Karkkainen, T. 2018. ``Price Discovery in the bitcoin Futures and Cash Markets." SSRN. Retrieved on the 15.10.2018 from \url{https://papers.ssrn.com/sol3/papers.cfm?abstract_id=3243969}.\vspace{0.2cm}
		
		Keister, T.,and D. Sanches. 2018. ``Managing Aggregate Liquidity: The Role of a Central
Bank Digital Currency." Semanticscholar. Retrieved on the 15.11.2018 from \url{https://www.semanticscholar.org/paper/Managing-Aggregate-Liquidity-%3A-The-Role-of-a-Bank-%E2%88%97-Keister-Sanches/feb3a1b19ea9b7915d5ede69fc75d77e5bf81d99}.\vspace{0.2cm}

		Kim, A., S. Trimborn, and W. H\"ardle. 2019. ``VCRIX - volatility index for crypto-currencies on the basis of CRIX." Forthcoming.\vspace{0.2cm}
		
		King, S., and S. Nadal. 2012. ``PPCoin: Peer-to-Peer Crypto-Currency with Proof-of-Stake." Retrieved on the 24.11.2018 from \url{https://decred.org/research/king2012.pdf}.\vspace{0.2cm}

		Klein, T., P. Hien, and T. Walther. 2018. ``bitcoin Is Not the New Gold: A Comparison of Volatility, Correlation, and Portfolio Performance." Retrieved on the 25.03.2018 from \url{http://dx.doi.org/10.2139/ssrn.3146845}.\vspace{0.2cm}

		Krueckeberg, S., and P. Scholz. 2018. ``Decentralized Efficiency? Arbitrage in bitcoin Markets".  http://dx.doi.org/10.2139/ssrn.3292127.\vspace{0.2cm}
		
		Lamport, L., R. Shostak, M. Pease. 1982. ``The Byzantine Generals Problem." ACM Transactions on Programming Languages and Systems. 4 (3): 382–401. doi:10.1145/357172.357176. \vspace{0.2cm}
		
		Lee, S. 2018. ``Bitcoin's Energy Consumption Can Power An Entire Country -- But EOS Is Trying To Fix That." Forbes. Retrieved on the 14.01.2019 from \url{https://www.forbes.com/sites/shermanlee/2018/04/19/bitcoins-energy-consumption-can-power-an-entire-country-but-eos-is-trying-to-fix-that/#79121fd01bc8}.\vspace{0.2cm}

		Liu, Y., Tsyvinski, A., and X. Wu. 2019a. ``Common Risk Factors in Cryptocurrency." SSRN. Retrieved on the 15.07.2019 from \url{https://papers.ssrn.com/sol3/papers.cfm?abstract_id=3379131}.\vspace{0.2cm}

		Liu, Y., Sheng, J, and W. Wang. 2019b. ``How Much Tech in FinTech? Evidence from Initial Coin Offerings." Working Paper. University of California Irvine.\vspace{0.2cm}
		
		Makarov, I., and A. Schoar. 2018. ``Trading and Arbitrage in Cryptocurrency Markets." The Journal of Financial Economics. Forthcoming.\vspace{0.2cm}
				
		MAS. 2017. ``Project Ubin: Central Bank Digital Money using Distributed Ledger Technology." Monetary Authority of Singapore. Retrieved on the 14.12.2018 from \url{http://www.mas.gov.sg/Singapore-Financial-Centre/Smart-Financial-Centre/Project-Ubin.aspx}.\vspace{0.2cm}
		
		Mishra, S., V. Jacob, and S. Radhakrishnan. 2018. ``Energy Consumption – Bitcoin’s Achilles Heel." Retrieved on the 14.01.2019 from \url{https://ssrn.com/abstract=3076734 or http://dx.doi.org/10.2139/ssrn.3076734}.\vspace{0.2cm}
		
		Momtaz, P. 2018. ``Initial Coin Offerings." SSRN. Retrieved on the 12.12.2018 from \url{https://ssrn.com/abstract=3166709}.\vspace{0.2cm}
		
		Nasekin, S., and C. YH. Chen. 2019. ``Deep Learning-Based Cryptocurrency Sentiment Construction." Digital Finance. Smart Data Analytics, Investment Innovation, and Financial Technology. ISSN 2524-6186. Forthcoming.\vspace{0.2cm}
		
		Nakamoto, S. 2008. ``Bitcoin: A Peer-to-Peer Electronic Cash System." Retrieved on the 14.12.2018 from \url{https://bitcoin.org/bitcoin.pdf}.\vspace{0.2cm}

Onatski, A., and C. Wang. 2018. ``Alternative Asymptotics for Cointegration Tests in Large VARs." Econometrica, Volume 86, Issue 4. P.p 1465-1478. https://doi.org/10.3982/ECTA14649.\vspace{0.2cm}
		
		Ong, B., L. Guo, and K. Lee. 2015. ``Evaluating the Potential of Alternative Cryptocurrencies." Elsevier, Academic Press. Pp. 81–135.\vspace{0.2cm}
		
		Paar, C., and J. Pelzl. 2010. ``Understanding Cryptography." Berlin: Springer‐Verlag.\vspace{0.2cm}
		
		 Panwar, G., Misra, S., and R. Vishwanathan. 2019. ``BlAnC: Blockchain-based Anonymous and Decentralized Credit Networks." Cryptology ePrint Archive: Report 2019/014. DOI: 10.1145/3292006.3300034.\vspace{0.2cm}
		
		Pagnotto, E., and A. Buraschi. 2018. ``An Equilibrium Valuation of Bitcoin and Decentralized Network Assets." SSRN.  Retrieved on the 20.01.2019 from \url{https://papers.ssrn.com/sol3/papers.cfm?abstract_id=3142022}.\vspace{0.2cm}

		Park, S., K. Pietrzak, J. Alwen, G. Fuchsbauer, and P. Gazi. 2015. ``Spacecoin: A cryptocurrency based on proofs of space." Cryptology ePrint Archive. Retrieved on the 24.11.2018 from \url{https://eprint.iacr.org/2015/528/20150623:221402}.\vspace{0.2cm}
		
		Petukhina, A., Trimborn, S., H\"ardle, W., and H. Elendner. 2018. ``Investing With Cryptocurrencies – Evaluating the Potential of Portfolio Allocation Strategies". SSRN. Retrieved on the 12.01.2019 from \url{https://papers.ssrn.com/sol3/papers.cfm?abstract_id=3274193}.\vspace{0.2cm}
		
			Pieters, G., and S. Vivianco. 2015. ``Bitcoin arbitrage and unofficial exchange rates." Trinity University. Retrieved on the 10.03.2019 from \url{http://www.ginapieters.com/uploads/1/0/2/8/10288430/bitcoin.pdf}.\vspace{0.2cm}
		
		Pop, C., T. Cioara, M. Antal, I. Anghel, I. Salomie, and M. Bertoncini. 2018. ``Blockchain Based Decentralized Management of Demand Response Programs in Smart Energy Grids." MDPI. Sensors 2018, 18, 162; doi:10.3390/s18010162.\vspace{0.2cm}
		
		Raskin, M., and D. Yermack. 2016. ``Digital Currencies, Decentralized Ledgers, and the Future of Central Banking". National Bureau of Economic Research. NBER Working Paper No. 22238.\vspace{0.2cm}

 Robinson, P., Hyland-Wood, D., Saltini, R., Johnson, S., and J. Brainard. 2019. ``Atomic Crosschain Transactions for Ethereum Private Sidechains." Retrieved on the 12.06.2019 from \url{https://arxiv.org/pdf/1904.12079.pdf}.\vspace{0.2cm}

		Santo, A., Minowa, I., Hosaka, G., Hayakawa, S., Kondo, M., Ichiki, S., and Y. Kaneko. 2016. Applicability of Distributed Ledger Technology
		to Capital Market Infrastructure. JPX Working Paper. Japan Exchange Group. Retrieved on the 12.12.2018 from \url{https://www.cnbc.com/2017/10/13/bitcoin-get-serious-about-digital-currency-imf-christine-lagarde-says.html}.\vspace{0.2cm}

		SARB. 2018. ``FinTech Programme - Media statement, Tuesday, 13 February 2018." South African Reserve Bank. Retrieved on the 15.12.2018 from\\
		\brokenurl{https://www.cnbc.com/2017/10/13/bitcoin-get-serious-about-digital-currency}{-imf-christine-lagarde-says.html}.\vspace{0.2cm}
		
		Scaillet, O., A. Treccani, and C. Trevisan. 2018. ``High-Frequency Jump Analysis of the bitcoin Market." Journal of Financial Econometrics. Forthcoming.\vspace{0.2cm}
		
		Schilling, L., and H. Uhlig. 2018. ``Some simple Bitcoin Economics". Becker Friedman Institute for Research in Economics Working Paper No. 2018-21.\vspace{0.2cm}

		Schuster, B. 2017. ``Ripple – Why You Shouldn’t Invest (and Not Because It’s a Scam)." Retrieved on the 12.12.2018 from \url{http://hivergent.com/you-shouldnt-invest-in-ripple-and-not-because-its-a-scam/}.\vspace{0.2cm}

		SEC. 2017a. SEC Issues Investigative Report Concluding DAO Tokens, a Digital Asset, Were Securities. U.S. Securities and Exchange Commission. Retrieved on the 14.12.2018 from \url{https://www.sec.gov/news/press-release/2017-131}.\vspace{0.2cm}
		
		SEC. 2017b. Investor Bulletin: Initial Coin Offerings. Retrieved on the 28.02.2018 from \url{https://www.investor.gov/additional-resources/news-alerts/alerts-bulletins/investor-bulletin-initial-coin-offerings}.\vspace{0.2cm}

		SEC. 2019. MEMORANDUM: Meeting with Bitwise Asset Management, Inc., NYSE Arca, Inc., and Vedder	Price P.C. Retrieved on the 23.03.2019 from \url{https://www.sec.gov/comments/sr-nysearca-2019-01/srnysearca201901-5164833-183434.pdf}.\vspace{0.2cm}

		Sharma, R. 2018. ``The Number Of Dark Pools In Cryptocurrency Trading Is Increasing." Investopedia. Retrieved on the 14.10.2018 from \url{https://www.investopedia.com/news/number-dark-pools-cryptocurrency-trading-increasing/}.\vspace{0.2cm}

		Shilov, K. 2018. ``Top 10 Jurisdictions for a Worry-Free ICO in the Current Regulatory Climate." Hackernoon. Retrieved on the 23.11.2018 from \url{https://hackernoon.com/top-10-jurisdictions-for-a-worry-free-ico-in-the-current-regulatory-climate-40bb04661454}.\vspace{0.2cm}

		Shu, S. and Z. Zhu. 2019. ``Real-time Prediction of Bitcoin Bubble Crashes." arXiv. Retrieved on the 28.05.2019 from \url{https://arxiv.org/abs/1905.09647}.\vspace{0.2cm}

		Sockin, M., and W. Xiong. 2018. ``A Model of Cryptocurrencies." Princeton University. Retrieved on the 14.02.2019 from \url{http://wxiong.mycpanel.princeton.edu/papers/Crypto.pdf}.\vspace{0.2cm}

		Taskinsoy, J. 2019. ``Facebook’s Project Libra: Will Libra Sputter Out or Spur Central Banks to Introduce Their Own Unique Cryptocurrency Projects?" Retrieved on the 26.07.2019 from \\
		\url{https://ssrn.com/abstract=3423453} .\vspace{0.2cm}
		
		Trimborn, S., and W. H\"ardle. 2018. ``CRIX an Index for cryptocurrencies." Journal of Empirical Finance.  https://doi.org/10.1016/j.jempfin.2018.08.004 .\vspace{0.2cm}
		
		Wildi, M., and N. Bundi. 2018. ``bitcoin and Market-(In)Efficiency: a Systematic Time
		Series Approach." Special Issue of Digital Finance on Cryptocurrencies. Digital Finance. Smart Data Analytics, Investment Innovation, and Financial Technology. ISSN 2524-6186.\vspace{0.2cm}

		World Bank. 2017. ``UFA2020 Overview: Universal Financial Access by 2020." The World Bank. Retrieved on the 26.04.2018 from \\
		\url{http://www.worldbank.org/en/topic/financialinclusion/brief/achieving-universal-financial-access-by-2020} .\vspace{0.2cm}

	\end{footnotesize}

\end{document}